\documentclass[aps, twocolumn, showpacs, letterpaper]{revtex4}

\usepackage{amsmath}
\usepackage{amssymb}
\usepackage{mathrsfs}
\usepackage{xspace}
\usepackage{graphicx}
\usepackage{bm}

\usepackage{ifpdf}
\ifpdf
\fi

\newcommand{\eq}[1]{(\ref{#1})}
\newcommand{\Eq}[1]{Eq.~(\ref{#1})}
\newcommand{\Eqs}[1]{Eqs.~(\ref{#1})}
\newcommand{\Fig}[1]{Fig.~\ref{#1}}
\newcommand{\Sec}[1]{Sec.~\ref{#1}}
\newcommand{\Ref}[1]{Ref.~\cite{#1}}
\newcommand{\Refs}[1]{Refs.~\cite{#1}}
\newcommand{\App}[1]{Appendix~\ref{#1}}

\newcommand{\eg}{{e.g.,\/}\xspace}
\newcommand{\ie}{{i.e.,\/}\xspace}

\renewcommand{\vec}[1]{{\mathbf{#1}}}
\newcommand{\pd}{\partial}

\newcommand{\mc}[1]{\mathcal{#1}}
\newcommand{\mcc}[1]{\mathfrak{#1}}
\newcommand{\msf}[1]{\mathsf{#1}}
\newcommand{\favr}[1]{\langle #1 \rangle}
\newcommand{\ffavr}[1]{\favr{\!\favr{#1}\!}}

\begin{document}

\title{Nonlinear frequency shift of electrostatic waves in general collisionless plasma:\\unifying theory of fluid and kinetic nonlinearities}

\author{Chang Liu and Ilya~Y. Dodin}

\affiliation{Department of Astrophysical Sciences, Princeton University, Princeton, New Jersey 08544, USA} 

\date{\today}


% 52.35.Mw - Nonlinear phenomena: waves, wave propagation, and other interactions 
%           (including parametric effects, mode coupling, ponderomotive effects, etc.) 
% 52.35.Sb - Solitons; BGK modes 
% 52.35.Fp - Electrostatic waves and oscillations
% 45.20.Jj - Lagrangian and Hamiltonian mechanics 

\begin{abstract}
The nonlinear frequency shift is derived in a transparent asymptotic form for intense Langmuir waves in general collisionless plasma. The formula describes both fluid and kinetic effects simultaneously. The fluid nonlinearity is expressed, for the first time, through the plasma dielectric function, and the kinetic nonlinearity accounts for both smooth distributions and trapped-particle beams. Various known limiting scalings are reproduced as special cases. The calculation avoids differential equations and can be extended straightforwardly to other nonlinear plasma waves. 
\end{abstract}

\maketitle

\bibliographystyle{brief} 

%%%%%%%%%%%%%%%%%%%%%%%%%%%%%%%%%%%%%%% 

\section{Introduction}

It has long been known that the nonlinear frequency shifts $\delta \omega$ of collisionless plasma waves can depend on the wave amplitude $a$ in various ways, even at small $a$. The contributions to $\delta\omega$ that stem from fluid effects typically scale as $\propto a^2$ \cite{ref:akhiezer56, ref:tidman65, ref:bertrand69, ref:coffey71, ref:dewar72e, ref:mckinstrie87}, whereas kinetic effects can produce contributions scaling as $\propto \sqrt{a}$ \cite{ref:manheimer71a, ref:morales72, ref:lee72, ref:dewar72b, ref:kim76, ref:rose01, ref:barnes04, ref:rose05, ref:lindberg07, ref:khain07, ref:benisti07, ref:benisti08, ref:benisti09, ref:matveev09, ref:schamel12} or even decreasing with $a$ \cite{ref:goldman71, ref:krasovsky94, ref:krasovsky92, ref:krasovskii95, ref:krasovsky07, ref:breizman10, my:trcomp}. The interest to such nonlinear dispersion relations (NDRs) has been revived recently in connection with laser-plasma interactions \cite{foot:berger13} and the evolution of energetic-particle modes in tokamaks \cite{foot:breizman11}. However, each of the existing theories that offers an explicit formula for $\delta \omega$ describes just one type of nonlinearity. The possible effect of multiple coexisting nonlinearities that yield different scalings in $a$ was explored in \Ref{ref:winjum07}, but only heuristically and for an artificial plasma model. The more general existing theories, such as that of Bernstein-Greene-Kruskal (BGK) waves \cite{ref:bernstein57}, rely on formal solutions of the Vlasov-Maxwell equations and are not always easy to apply in practice due to their complexity. But is it possible to derive a comprehensive and yet tractable asymptotic NDR for general collisionless plasma? 

The answer is yes, and here we show how to do it explicitly. We build on the theory that was developed recently in \Refs{my:itervar, my:bgk, my:acti, my:actii, my:actiii} and represents a fully nonlinear kinetic version of Whitham's average-Lagrangian approach \cite{book:whitham, foot:khain}. (We assume that, when present, resonant particles are phase-mixed, so the corresponding waves are of the BGK type \cite{foot:diss}.) This allows deriving the NDR directly from the wave Lagrangian, which is known, without solving any differential equations. In fact, all wave properties can be traced to the properties of a single function characterizing individual particles, namely, the normalized action $j$ of a particle as a function of its normalized energy $r$. We focus on Langmuir waves in one-dimensional electron plasma, but extending the theory to general waves is straightforward to do.

Unlike in \Refs{my:bgk, my:acti, my:actii, my:actiii}, where a related theory was constructed under the sinusoidal-wave approximation, we now allow for a nonzero amplitude of the second harmonic (higher-order harmonics are assumed negligible) and find this amplitude self-consistently. Kinetic and fluid nonlinearities are hence treated on the same footing, and two types of distributions are studied in detail: (i) First, we consider particle distributions, $f_0$, that are smooth enough near the resonance. We assume that $a$ is small in this case; then kinetic nonlinearities scale as half-integer powers of $a$, and fluid nonlinearities scale as integer powers of $a$. In particular, it is shown for the first time that the fluid nonlinearity can be expressed in terms of the plasma dielectric function and, contrary to the common presumption, can have either sign. (ii) Next, we consider the effect of abrupt trapped-particle beam distributions superimposed on smooth $f_0$. The small parameter in this case is not $a$ but rather the appropriately normalized trapped-particle density. We show that such beams contribute additively to the NDR. We revisit the case of flat-top trapped beams \cite{ref:krasovsky07, ref:breizman10}, as an illustration, and demonstrate that our asymptotic theory yields predictions for both the wave frequency and the second-harmonic amplitude that are virtually indistinguishable from those given by the exact solution of the Vlasov-Poisson system. Our theory then can be considered as advantageous over such solutions. This is because, while offering a reasonable precision, it is more transparent and flexible, \ie can be used also when exact analytical formulas do not exist.

Nonlinearities with both smooth particle distribution (including fluid and kinetic effects) and distributions with abrupt changes near the resonant region (beam distribution) are treated on the same footing.  In particular, we show, for the first time, that the fluid nonlinearity can be expressed in terms of the plasma dielectric function and, contrary to the common presumption, can have either sign. Nonlinearities due to abrupt distributions of trapped beams are described separately and can be added to the fluid and kinetic nonlinearities.

The paper is organized as follows. In \Sec{sec:basic}, we briefly review the underlying variational approach and introduce basic notation. In \Sec{sec:model}, we describe the wave model and asymptotics of some auxiliary functions derived from $j(r)$. In \Sec{sec:smooth}, we derive the general expression for $\delta \omega$ for smooth $f_0$ and present examples, including the cases of cold, waterbag, kappa, and Maxwellian $f_0$. In \Sec{sec:trapped}, we discuss the effect of trapped-particle beams superimposed on a smooth $f_0$. In \Sec{sec:conc}, we summarize the main results of our work. Some auxiliary calculations are also presented in appendixes.

%%%%%%%%%%%%%%%%%%%%%%%%%%%%%%%%%%%%%%% 
\section{Basic concepts and notation}
\label{sec:basic}

%--------------------------------------
\subsection{Wave Lagrangian density}
\label{sec:wld}

As reviewed in \Refs{my:itervar}, the dynamics of an adiabatic plasma wave can be derived from the least action principle, $\delta \Lambda = 0$, where $\Lambda$ is the action integral given by
\begin{gather}
\Lambda =\int \Big[\favr{\mcc{L}_{\rm em}} - \sum_{\msf{s}} \bar{n}_{\msf{s}} \ffavr{\mc{H}_{\msf{s}}}\Big]\,dt\,d^3x.
\end{gather}
Here $\mcc{L}_{\rm em}$ is the Lagrangian density of electromagnetic field in vacuum, $\mc{H}_{\msf{s}}$ are the oscillation-center (OC) Hamiltonians of single particles of type $\mathsf{s}$, the summation is taken over all particle types, $\bar{n}_{\msf{s}}$ are the corresponding average densities, $\favr{\ldots}$ denotes averaging over rapid oscillations in time and space, and $\ffavr{\ldots}$ denotes averaging over the local distributions of the particle OC canonical momenta. Keep in mind that these distributions must be treated as prescribed when the least action principle is applied to derive the NDR. 

We will limit our consideration to electrostatic oscillations in one-dimensional nonrelativistic electron plasma. In this case, $\mcc{L}_{\rm em} = E^2/8\pi$, where $E = - \pd_x \phi$ is the electric field, and $\phi$ is the potential. We will allow $\phi$ to consist of a rapidly oscillating potential $\tilde{\phi}$ and, possibly, a potential $\bar{\phi}$ that is ``low-frequency'' (LF) both in time and space. Depending on a specific problem (\App{app:E}), the LF field, $\bar{E} \doteq - \pd_x\bar{\phi}$, may be zero, determined externally, or result from wave dynamics, in which case it can also be calculated, if needed. However, \textit{finding $\bar{E}$ is a problem that is separate from finding the NDR}. This is because, strictly speaking, a local NDR is well defined only in the (possibly noninertial) frame of reference $\mc{K}$ where the $\bar{E}$-driven acceleration vanishes, at least locally \cite{foot:static}. Denoting the new coordinate as $\hat{x}$, one can write
\begin{gather}
\Lambda = \int \mcc{L}\,\mc{J}[\bar{E}]\,dt\,d^3\hat{x}.
\end{gather}
Here,  the wave Lagrangian density, $\mcc{L}$, is given by
\begin{gather}
\mcc{L} = \frac{(\pd_x \bar{E})^2}{8\pi} + \frac{\favr{(\pd_x \tilde{\phi})^2}}{8\pi} - \sum_{\msf{s}} \bar{n}_{\msf{s}} \ffavr{\mc{H}_{\msf{s}}}\label{eq:lagrangian},
\end{gather}
and $\mc{J}[\bar{E}]$ is the Jacobian of the coordinate transformation. Since the mapping $x \mapsto \hat{x}$ is independent of $\tilde{\phi}$, the fact that $\mc{J}$ is generally a functional of $\bar{E}$ has no effect on the NDR \cite{foot:clear}. We can also omit the term $(\pd_x \bar{\phi})^2/(8\pi)$ in \Eq{eq:lagrangian} for the same reason.

It is thereby seen that whether the LF field is zero or not is irrelevant to further calculations. (Switching between $\mc{K}$ and the inertial frame, or remapping $x \leftrightarrow \hat{x}$, merely introduces or removes a chirp determined by $\bar{E}$.) Hence we will call $\mc{K}$ the laboratory frame for simplicity. In case of a periodic system, $\mc{K}$ can be the true laboratory system indeed (\App{app:E}). Otherwise, one can always choose $\mc{K}$ such that, locally, its velocity relative to the true laboratory frame be zero, even though the relative acceleration may be nonzero. Also keep in mind that the laboratory frame of reference is not necessarily the plasma rest frame.

%--------------------------------------
\subsection{OC Hamiltonians}

The OC Hamiltonians of passing particles (denoted with index $\msf{p}$) and trapped particles (denoted with index $\msf{t}$), which enter \Eq{eq:lagrangian}, can be expressed as follows:
\begin{gather}\label{eq:hpht}
\mc{H}_{\msf{p}} = Pu + \mc{E} - mu^2/2,
\quad 
\mc{H}_{\msf{t}} = \mc{E} - mu^2/2.
\end{gather}
Here $u \doteq \omega/k$ is the wave phase velocity (we use the symbol $\doteq$ for definitions), $\omega \doteq - \pd_t \xi$ is the local frequency, $k \doteq \pd_x \xi$ is the local wave number, and $\xi$ is the wave phase. Also, $P$ is the OC canonical momentum of a passing particle in the laboratory frame $\mc{K}$, 
\begin{gather}\label{eq:mcE}
\mc{E} \doteq m w^2/2 + e\phi
\end{gather}
is the particle total energy in the frame $\mc{K}'$ that moves with respect to $\mc{K}$ at velocity $u$, $w$ is the particle velocity in $\mc{K}'$, and $m$ and $e$ are the electron mass and charge. Notice also, for future references, that, at small enough $\phi$, $\mc{H}_{\msf{p}}$ becomes the well known ponderomotive Hamiltonian,
\begin{gather}
\mc{H}_{\msf{p}} \approx \frac{P^2}{2m} + \Phi.\label{eq:Hpond}
\end{gather}
Here the ``ponderomotive potential'' $\Phi$ is given by
\begin{gather}\label{eq:HPhi}
\Phi = \frac{e^2 k^2 \phi_1^2}{4m(\omega - kP/m)^2},
\end{gather}
and $\phi_1$ is the amplitude of $\tilde{\phi}$. The OC canonical momentum $P$ can then be expressed as
\begin{gather}\label{eq:PPhi}
P \approx mV - \pd_V \Phi,
\end{gather}
where $V$ is the average velocity, or the OC velocity.

\begin{figure}
\centering
\includegraphics[width=.48\textwidth]{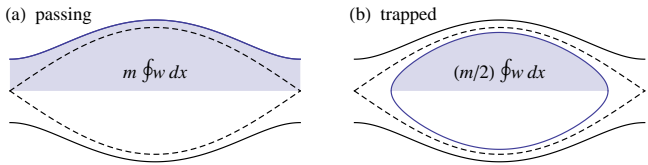}
\caption{Schematic of particle trajectories in phase space, illustrating the definition of $2 \pi J$ (shaded area): (a) for a passing particle, (b) for a trapped particle. For $J$ to be continuous at the separatrix (dashed), with the passing-particle action defined as $2\pi J = m \oint w\,dx$, for trapped particles one must use the definition $2\pi J = (m/2) \oint w\,dx$. The figure is an adjustment of Fig.~1 from \Ref{my:actii}.}
\label{fig:J}
\end{figure}

Using \Eqs{eq:hpht}, one can cast \Eq{eq:lagrangian} as
\begin{gather}\label{eq:L2}
\mcc{L} =  \frac{\favr{(\pd_x \tilde{\phi})^2}}{8\pi} - \bar{n} \ffavr{\mc{E}} - \bar{n}_{\msf{p}} \ffavr{P} u + \frac{\bar{n}mu^2}{2},
\end{gather}
where $\bar{n} \doteq \bar{n}_{\msf{p}} + \bar{n}_{\msf{t}}$ is the total average density. Let us also introduce $J$ as the particle action variable in $\mc{K}'$, \ie the (appropriately normalized) phase space area that is swept by the particle trajectory on a single period, $J \propto m \oint w\,dx$ (\Fig{fig:J}). The separatrix action, $J_*$, can be estimated as $J_* \sim \hat{J}\sqrt{a_p}$, where we introduced
\begin{gather}
\hat{J} \doteq \frac{m\omega}{k^2}, \quad a_p \doteq \frac{e k^2 \phi_1}{m \omega_p^2}.
\end{gather}
Trapped particles, for which $J$ is the OC canonical momentum, have $0 < J < J_*$. Passing particles, for which 
\begin{gather}\label{eq:P}
P = mu + kJ\,\text{sgn}(w),
\end{gather}
have $J > J_*$. One can then rewrite \Eq{eq:L2} as follows:
\begin{gather}\label{eq:L3}
\mcc{L} = \frac{\favr{(\pd_x \tilde{\phi})^2}}{8\pi} - \bar{n} \ffavr{\mc{E}} + \Delta \mcc{L}.
\end{gather}
Here $\Delta \mcc{L}$ is independent of the wave amplitude, so it has no effect on the NDR, as will become clear shortly \cite{foot:clear}. 

%--------------------------------------
\subsection{Action distribution}
\label{sec:distr}

The ensemble averaging can be expressed through the action distribution $F(J)$ that includes both trapped and passing particles, $F(J) = F_{\msf{t}}(J) + F_{\msf{p}}(J)$, where $F_{\msf{t},\msf{p}}(J \gtrless J_*) \equiv 0$. We will assume the normalization such that $\int^\infty_0 F_{\msf{t},\msf{p}}(J)\,dJ = \bar{n}_{\msf{t},\msf{p}}/\bar{n}$; then, $\int^\infty_0 F(J)\,dJ = 1$, and
\begin{gather}
\ffavr{\ldots} = \int^\infty_0 (\ldots)\, F(J)\,dJ.
\end{gather}

Note that the action distribution is not different from, say, the velocity distribution in the sense that, if needed, $F(J)$ can be calculated self-consistently \cite{foot:benisti14}. Yet $F(J)$ is also advantageous in the sense that it is \textit{easier} to calculate or model in many cases of interest, especially those that are of interest to approach analytically. This is because $J$ is \textit{the} natural variable for phase-mixed distributions. For example, $J$ is an adiabatic invariant for trapped particles, and passing particles often conserve $P$ or $\mc{H}_{\msf{p}}$, which are expressible through $J$. Models of the distribution functions in such variables are therefore particularly robust. Hence, the problem of modeling $F(J)$ often can be decoupled from the problem of finding the NDR. In other words, it can be meaningful to search for $\omega$ for a given $F(J)$. (The situation is similar to calculating linear dispersion for a given ``unperturbed'' velocity distribution \cite[Chap.~8]{book:stix}, but nonlinear calculations involving velocity distributions usually do not enjoy this property.) Apart from the examples considered in this paper, see also \Refs{my:actiii, my:nmi, my:trcomp} where such decoupling is utilized to arrive at results of practical interest.

Below we will present several models for $F(J)$ and demonstrate that they readily yield the many scalings for $\delta\omega$ that were reported previously for various special cases. We postpone considering beam distributions until \Sec{sec:trapped}, whereas the bulk plasma will be modeled as follows. Let us assume that a wave is excited by a driver with homogeneous amplitude, so the OC canonical momenta $P$ are conserved and equal the initial momenta,
\begin{gather}\label{eq:PmV0}
P = mV_0.
\end{gather}
From \Eq{eq:P}, one gets
\begin{gather}\label{eq:J0}
J = |1 - V_0/u|\hat{J},
\end{gather}
so the action distribution is given by
\begin{gather}\label{eq:Ff0}
F(J) = \frac{k}{m} \left[f_0\left(u + \frac{k J}{m}\right) + f_0\left(u - \frac{k J}{m}\right)\right],
\end{gather}
where $f_0$ is the initial velocity distribution. This is also known as the adiabatic-excitation model \cite{ref:dewar72b, foot:berger13}. 

%--------------------------------------
\subsection{Nonlinear Doppler shift}
\label{sec:doppler}

Before we proceed to a general calculation, notice the following. The conservation of $P$ implies that, as a result of the wave excitation, the plasma generally changes its average velocity, $\ffavr{V}$, by some $\ffavr{\Delta V}$. This leads to a nonlinear Doppler shift, $k\ffavr{\Delta V}$, by which the frequency $\omega$ in the laboratory frame $\mc{K}$ differs from that in the plasma rest frame \cite{ref:dewar72e, foot:addif}. The shift can be calculated using that $V_{\msf{t}} = u$ and $V_{\msf{p}} = \pd_P \mc{H}_{\msf{p}} = u + (\pd_J \mc{E})\, \text{sgn}(w)$, where $\mc{E}$ is considered as a function of $J$ and of the wave parameters \cite{my:itervar}. However, the concept of the plasma rest frame is meaningful mainly when the resonant population is negligible, in which case the following simple derivation is possible. (Also see \App{app:hydro} for an alternative derivation under the same condition.)

By combining \Eq{eq:PPhi} with \Eq{eq:PmV0}, one gets for an initially-resting plasma that
\begin{gather}\label{eq:VPhi}
\ffavr{\Delta V} = \ffavr{V - V_0} \approx \frac{\ffavr{\pd_V \Phi}}{m} \approx \frac{ua^2}{2}\,\ffavr{(1 - V_0/u)^{-3}}.
\end{gather}
One can also reexpress the right hand side as follows:
\begin{align}\notag
\frac{\ffavr{\pd_V \Phi}}{m} 
& \approx \frac{e^2 k^2 \phi_1^2}{4m^2}\,\int^\infty_{-\infty} [\pd_{V_0} (\omega - kV_0)^{-2}] f_0(V_0)\, dV_0 \notag \\
& \approx -\frac{e^2 k \phi_1^2}{4m^2}\,\frac{\pd}{\pd \omega}\int^\infty_{-\infty} \frac{f_0(V_0)}{(V_0-\omega/k)^2}\, dV_0 \notag \\
& = \frac{a^2}{4}\,\frac{\omega^4}{k\omega_p^2}\,\frac{\pd \epsilon(\omega, k)}{\pd \omega},
\end{align}
where
\begin{multline}\label{eq:eps0}
\epsilon(\omega, k) \doteq 1 - \frac{\omega_p^2}{k^2} \int^\infty_{-\infty} \frac{f_0(V_0)}{(V_0 - \omega/k)^2}\, dV_0
\\ = 1 - \frac{\omega_p^2}{k^2} \int^\infty_{-\infty} \frac{f_0'(V_0)}{V_0 - \omega/k}\, dV_0.
\end{multline}
This leads to
\begin{gather}\label{eq:ndop0}
\ffavr{\Delta V} = \frac{a^2}{4}\,\frac{\omega^4}{k\omega_p^2}\,\frac{\pd \epsilon(\omega, k)}{\pd \omega},
\end{gather}
where, within the adopted accuracy, the right hand side must be evaluated at the linear frequency, $\omega_0$. Hence, one eventually arrives at the following expression for the nonlinear Doppler shift:
\begin{gather}\label{eq:ndop}
k\ffavr{\Delta V} = \frac{a_0^2}{4}\,\frac{\omega_0^4}{\omega_p^2}\,\frac{\pd \epsilon(\omega_0, k)}{\pd \omega_0},
\end{gather}
where we introduced $a_0 \doteq e k^2 \phi_1/(m \omega_0^2)$. Also notably, the momentum density associated with the nonlinear Doppler shift is
\begin{gather}\label{eq:wmom}
m \bar{n} \ffavr{\Delta V} = \frac{k\tilde{E}^2}{16\pi}\,\frac{\pd \epsilon(\omega_0, k)}{\pd \omega_0} = \frac{k\mcc{E}}{\omega_0},
\end{gather}
where $\tilde{E} \doteq k\phi_1$ is the wave amplitude, and $\mcc{E}$ is the wave energy density \cite[Sec.~4-4]{book:stix}. One may recognize $k\mcc{E}/\omega_0$ as the canonical \cite{my:amc} momentum density of a linear wave \cite[Sec.~16-3]{book:stix}. It is seen then that the nonlinear Doppler shift is precisely the effect that allows a linear electrostatic wave carry momentum. (The additional acceleration caused by a LF field, if any, changes the momentum of the OC motion rather than the wave momentum.)

Some calculations reported in literature ignore the nonlinear Doppler shift and present the NDR as a relation between $\delta\omega$ and $a_0$ in the plasma rest frame \cite{foot:sagdeev}. Keep in mind, however, that this frame is by default unknown. Hence, calculating $\delta \omega$ in the plasma rest frame generally says nothing about the frequency shift that could be directly measured in laboratory. (Defining the wave momentum is also problematic in this case.) Although it is in principle possible to excite a wave without accelerating plasma, such scenarios are not typical for traveling waves that we consider \cite{ref:dewar72e}. Of course, there are geometries where the average flow is bound to vanish eventually, \eg to prevent the growth of the LF field (\App{app:E}). But this effect is not universal and, in any case, is restricted to stationary waves. In contrast, our Lagrangian formulation allows calculating a more fundamental, \textit{local} NDR, since $\bar{\phi}$ is treated as an independent function. As a result, our theory automatically accounts for plasma acceleration associated with the wave excitation in the frame $\mc{K}$ for a given particle distribution. (As we explained in \Sec{sec:wld}, the possibly nonzero simultaneous acceleration caused by $\bar{\phi}$ is not a part of the NDR and can be found separately, if needed.) Hence, $\delta\omega$ that we calculate below \textit{already includes} the nonlinear Doppler shift, and \Eq{eq:ndop} will be used only to establish connections between our results and other theories. 

%--------------------------------------
\subsection{Dielectric function}

It is to be noted that integration by parts in \Eq{eq:eps0} is justified only by the assumption of having no resonant particles, in which case the integrand is analytic. More generally, we will define $\epsilon$ as the following real function,
\begin{gather}\label{eq:epsP}
\epsilon(\omega, k) \doteq 1 - \frac{\omega_p^2}{k^2}\,\msf{P}\int_{-\infty}^\infty \frac{f_0'(v)}{v-\omega/k}\,dw,
\end{gather}
where $\msf{P}$ denotes the Cauchy principal value. [In the absence of resonant particles, \Eq{eq:epsP} is equivalent to \Eq{eq:eps0}.] For the adiabatic waves of interest, which are by definition phase-mixed and exhibit no Landau damping, such $\epsilon$ can be recognized as the linear dielectric function \cite[Chap.~8]{book:stix}.

%%%%%%%%%%%%%%%%%%%%%%%%%%%%%%%%%%%%%%% 
\section{Wave model}
\label{sec:model}

%--------------------------------------
\subsection{General dispersion relation}
\label{sec:gdr}

Since large amplitudes typically result in rapid deterioration of waves through various nonlinear instabilities, the very concept of the NDR is of interest mainly for waves with small enough amplitudes, \ie when the wave shape is close to sinusoidal. Harmonics of order $\ell > 1$ (which are phase-locked to the fundamental harmonic, $\ell = 1$) can then be treated as perturbations, and the spectrum decreases rapidly with $\ell$ \cite{foot:keen}. Typical calculations of the trapped-particle nonlinearities, such as those reviewed in \Ref{my:actii}, are restricted to the sinusoidal wave model, which neglects all harmonics with $\ell > 1$. Below, we extend those calculations by introducing a ``bisinusoidal'' wave model, which retains the first \textit{and} second harmonic, while neglecting those with $\ell > 2$.

Within the bisinusoidal wave model, one can search for the wave electrostatic potential in the form
\begin{gather}
\tilde{\phi} = \phi_1 \cos(\xi) + \phi_2 \cos(2\xi + \chi),
\end{gather}
where $\phi_1$, $\phi_2$, and $\chi$ are slow functions of $(t, x)$. (We will assume that $\phi_2$ is small enough, so there is only one minimum of the potential energy per wavelength; see below.) It is then convenient to adopt $\xi$, $\chi$, and
\begin{gather}
a_1 \doteq e\phi_1 k^2/(m\omega^2), \quad a_2 \doteq e\phi_2 k^2/(m\omega^2)
\end{gather}
as four independent variables. Minimizing $\Lambda$ with respect to $\xi$ leads to a dynamic equation for the wave amplitude \cite{my:itervar}, which is not of interest in the context of this paper. The remaining Euler-Lagrange equations are as follows: 
\begin{gather}
\pd_{a_1}\mcc{L} = 0, \quad \pd_{a_2}\mcc{L} = 0, \quad \pd_\chi \mcc{L} = 0,
\end{gather}
where $\mcc{L} = \mcc{L}(a_1, a_2, \chi, \omega, k)$. By substituting
\begin{gather}
\favr{E^2} = \Big(\frac{a_1^2}{2} + 2a_2^2\Big)\Big(\frac{m \omega u}{e}\Big)^2,
\end{gather}
one hence obtains the following three equations,
\begin{align}
 0 = & \frac{\pd\ffavr{\mc{E}}}{\pd a_1} - \frac{a_1}{8\pi \bar{n}}\left(\frac{m \omega u}{e}\right)^2,\label{eq:aux1}\\
 0 = & \frac{\pd\ffavr{\mc{E}}}{\pd a_2} - \frac{a_2}{2\pi \bar{n}}\left(\frac{m \omega u}{e}\right)^2,\label{eq:aux2}\\
 0 = & \frac{\pd\ffavr{\mc{E}}}{\pd \chi},\label{eq:chimain}
\end{align}
where $\mc{E} = \mc{E}(J, a_1, a_2, \chi, \omega, k)$. After eliminating $a_2$ and $\chi$ and introducing $a \doteq a_1$, one arrives at a single equation for $\omega(k, a)$, which constitutes the NDR.

%--------------------------------------
\subsection{Eliminating $\boldsymbol{\chi}$}
\label{sec:elchi}

Our first step is to eliminate $\chi$. To do this, consider \Eq{eq:chimain} in the following form:
\begin{gather}
\ffavr{\pd_\chi \mc{E}(J, a_1, a_2, \chi, \omega, k)} = 0. \label{eq:chi3}
\end{gather}
In order to express $\mc{E}$ as a function of $J$, we will need an explicit formula for $J$. A universal formula that applies to both trapped and passing particles is \cite{my:itervar}
\begin{gather}\label{eq:aux11}
J = \text{Re} \int^{2\pi/k}_0 \sqrt{2m[\mc{E} - e \phi(x)]}\,\frac{dx}{2\pi}.
\end{gather}
Using the notation
\begin{gather}
z \doteq \frac{a_2}{2a_1}, \quad r \doteq \frac{1}{2}\left(\frac{\mc{E}}{mu^2 a} + 1\right), 
\end{gather}
it is convenient to cast \Eq{eq:aux11} as
\begin{gather}\notag
J = \frac{\hat{J}\sqrt{a}}{\pi}\, \text{Re} \int^{\pi}_{- \pi} \left[r - \sin^2(\theta/2) - z \cos(2\theta + \chi)\right]^{1/2} d\theta.
\end{gather}
Hence we can rewrite \Eq{eq:chi3} as follows \cite{foot:dJ}:
\begin{gather}\label{eq:chi4}
\ffavr{\pd_\chi J/\pd_r J} = 0,
\end{gather}
where $J = J(r, a, z, \chi, \omega, k)$. Notice now that 
\begin{multline}\notag
(\pd_\chi J)|_{\chi = 0} = \frac{\hat{J}\sqrt{a}}{\pi}\, \text{Re} \int^{\pi}_{- \pi} d\theta \\ 
\times z\sin(2\theta)\left[r - \sin^2(\theta/2) - z \cos(2\theta)\right]^{-1/2},
\end{multline}
which is zero, because the integrand is an odd function of $\theta$. Therefore, \Eq{eq:chi4} has an obvious solution, $\chi = 0$ (or $\chi = \pi$, but the difference between these solutions can be absorbed in the sign of $z$). We will assume this solution without searching for others, because it corresponds to what seems to be the only (relatively) stable equilibrium seen in simulations, even for large-amplitude waves \cite{foot:sym}. For an alternative argument, see \cite{foot:z}.

The remaining equations that constitute the NDR [\Eqs{eq:aux1} and \eq{eq:aux2}] can then be represented as follows,
\begin{gather}
\mc{G}_1 - \frac{a_1\omega^2}{2\omega_p^2} = 0, \quad \mc{G}_2 - \frac{2 a_2 \omega^2}{\omega_p^2} = 0,
\end{gather}
or, equivalently,
\begin{gather}
1 - 2\mc{G}_1/a_p = 0, \quad 8z \mc{G}_1 - \mc{G}_2 = 0.\label{eq:ndr2}
\end{gather}
Here $\omega_p \doteq (4 \pi \bar{n} e^2/m)^{1/2}$, $\mc{G}_{1,2} \doteq \ffavr{G_{1,2}}$, and
\begin{gather}\label{eq:G12}
G_{1,2} \doteq \frac{\pd}{\pd a_{1,2}}\left[\frac{\mc{E}(J, a_1, a_2, \omega, k)}{mu^2}\right].
\end{gather}
Notably, $\pd_{a_2} G_1(J, a_1, a_2, \omega, k) = \pd_{a_1} G_2(J, a_1, a_2, \omega, k)$ by definition, and the same applies to $\mc{G}_{1,2}$. Also notably, $G_{1,2}$ have a clear physical meaning, which is as follows. The function $2G_1/a$ can be understood as the relative contribution of a particle with given action (or energy) to the wave squared frequency, $\omega^2$, in units $\omega_p^2$. The function $G_2/2$ can be understood as the particle relative contribution to the second harmonic amplitude, $\phi_2$, in units $m\omega_p^2/(ek^2)$ [because $a_2\omega^2/\omega_p^2 = e\phi_2k^2/(m\omega_p^2)$].

%--------------------------------------
\subsection{Normalized action}

\begin{figure*}[t]
\centering
\includegraphics[width=1\textwidth]{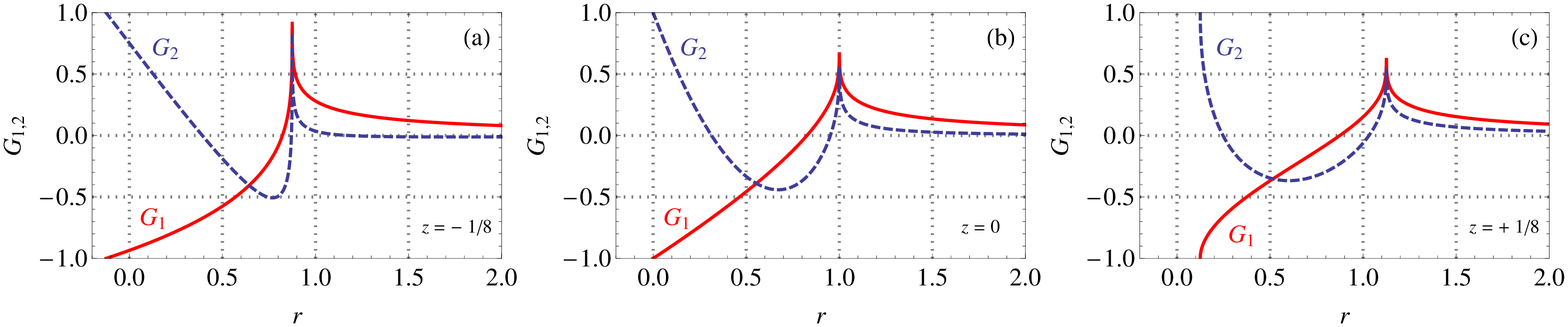}
\caption{(Color online) $G_1(r, z)$ (solid) and $G_2(r, z)$ (dashed) vs $r$ at fixed $z$: (a) $z = -1/8$, (b) $z = 0$, (c) $z = + 1/8$. Notice the singularity at $r = 1 + z$, which corresponds to the boundary between passing and trapped trajectories (\Sec{sec:refl}).}
\label{fig:g}
\end{figure*}

It is convenient to represent the particle action as $J = \hat{J}\sqrt{a}\,j(r, z)$, where the ``normalized action'', $j(r, z)$, is a dimensionless function given by
\begin{gather}
j(r, z) \doteq \frac{2}{\pi}\,\text{Re} \int^\pi_0 \sqrt{\mc{R}(r,z,\theta)}\,d\theta,\\
\mc{R}(r, z, \theta) \doteq r - \sin^2(\theta/2) - z \cos(2\theta).
\end{gather}
Then, $G_{1,2}$ can be expressed through $j$ alone,
\begin{align}
G_1 & = 2r - 1 + 2a_1\, \pd_{a_1} r(J, a_1, a_2, \omega, k) \notag\\
    & = 2r - 1 - 2a_1 \, \frac{\pd_{a_1} J(r, a_1, a_2, \omega, k)}{\pd_r J(r, a_1, a_2, \omega, k)} \notag \\
    & = 2r - 1 - \frac{j - 2z j_z}{j_r},\label{eq:Gdef}\\
G_2 & = 2a_1 \, \pd_{a_2} r(J, a_1, a_2, \omega, k) \notag\\
    & = - 2a_1 \, \frac{\pd_{a_2} J(r, a_1, a_2, \omega, k)}{\pd_r J(r, a_1, a_2, \omega, k)}\notag\\
    & = - \frac{j_z}{j_r},\label{eq:Kdef}
\end{align}
where $j \equiv j(r, z)$, and the indexes $r$ and $z$ denote partial derivatives. Explicit formulas for $j(r, z)$ and $G_{1,2}(r, z)$ are given in \App{app:J}. Characteristic plots of $G_{1,2}$ as functions of $r$ are presented in \Fig{fig:g}.

%--------------------------------------
\subsection{Restrictions on $\boldsymbol{r}$ and $\boldsymbol{z}$:\\reflection points and the trapping condition}
\label{sec:refl}

\begin{figure}[b]
\centering
\includegraphics[width=.48\textwidth]{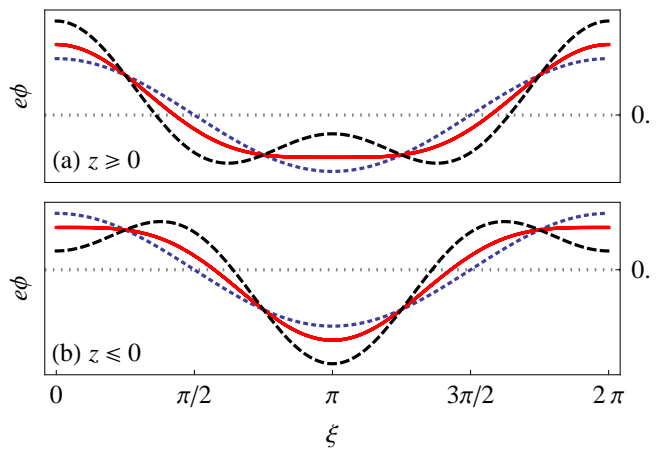}
\caption{(Color online) Potential energy, $e\phi$, vs the wave phase, $\xi$, for $z = 0$ (dotted blue), $|z| = 1/8$ (solid red), and $|z| = 1/3$ (dashed black): (a) $z \geqslant 0$, (b) $z \leqslant 0$. A second (per period) minimum of $e\phi$ appears at $|z| > 1/8$.}
\label{fig:phi}
\end{figure}

Let us also introduce an alternative, all-real representation of $j$,
\begin{gather}
j(r, z) = \frac{2}{\pi}\int^{\theta_0}_0 \sqrt{\mc{R}(r, z, \theta)}\, d\theta.
\end{gather}
Here, for passing particles $\theta_0$ equals $\pi$ and for trapped particles $\theta_0$ equals the first positive solution of
\begin{gather}\label{eq:g0}
\mc{R}(r, z, \theta_0) = 0.
\end{gather}
Equation \eq{eq:g0} can be expressed as the following quadratic equation for $y \doteq \cos \theta_0$,
\begin{gather}
4 z y^2 - y - 2r + 1 - 2z = 0.
\end{gather}
That gives $y = (1 \pm D)/(8z)$, where
\begin{gather}
D \doteq \sqrt{1 - 16 z + 32 r z + 32 z^2}.
\end{gather}
The solution for $y$ must be combined with the conditions
\begin{gather}
-1 \leqslant y \leqslant 1, \quad -1/8 \leqslant z \leqslant 1/8,
\end{gather}
which flow, respectively, from the definition of $y$ and from the assumption of having a single minimum of the potential energy per wavelength (\Fig{fig:phi}), which we adopted earlier. That leaves only one of the roots,
\begin{gather}
\theta_0 = \text{arccos}\left(\frac{1 - D}{8z}\right),
\end{gather}
and leads to the following trapping condition,
\begin{gather}
z \leqslant r < 1 + z,
\end{gather}
Notably, this condition also guarantees that $D$ and $\theta_0$ are real for trapped particles.

%--------------------------------------
\subsection{Asymptotics}

Below, we will also need asymptotics of $j$ and $G_{1,2}$. For deeply trapped particles (${r \to z}$), one can show that 
\begin{gather} 
j(r \to z) = 0, \quad 
G_{1,2}(r \to z) = \mp 1, \label{eq:glim0}
\end{gather}
whereas, for passing particles far from the resonance (${r \gg 1}$), the asymptotics are derived as follows \cite{foot:math}. Away from the separatrix, $j(r, z)$ is an analytic function of $z$. Assuming $z \ll 1$ (which is verified \textit{a~posteriori}), one can hence replace $j(r, z)$ with its Taylor expansion,
\begin{gather}
j(r, z) = \sum^\infty_{n = 0} z^n I_n(r). \label{eq:jexp0}
\end{gather}
The coefficients $I_n$ are calculated (\App{app:J}) using
\begin{gather}
I_n(r) = \frac{2}{\pi}\, \frac{(-1)^n}{n!}\frac{d^n}{dr^n}\int^\pi_0 \cos^n(2\theta)\,\sqrt{r - \sin^2(\theta/2)}\,d\theta.\notag
\end{gather}
By also expanding these functions in $1/r$, one gets
\begin{multline}
j(r, z) = \sqrt{r}\bigg[2 - \frac{1}{2r} - \frac{3 + 4z^2}{32 r^2}-\frac{5 + 3z + 12z^2}{128 r^3}\\ 
- \frac{5 \left(35+48 z+144 z^2\right)}{8192 r^4} \ldots \bigg],\label{eq:jexp}
\end{multline}
where we omitted terms of higher orders (in both $1/r$ and $z$) as insignificant for our purposes. The inverse series is then found to be
\begin{gather}\label{eq:rj}
r(j, z) = \frac{j^2}{4} + \frac{1}{2} + \frac{1 + 4z^2}{8j^2} + \frac{3z}{8j^4} + \frac{5(1 + 32 z^2)}{128 j^6} + \ldots
\end{gather}
This leads to the following equation for $\mc{E}$,
\begin{multline}
\frac{\mc{E}}{mu^2} = \frac{(J/\hat{J})^2}{2}+\frac{a_1^2+a_2^2}{4(J/\hat{J})^2}+\frac{3 a_1^2 a_2}{8 (J/\hat{J})^4}\\
+\frac{5(a_1^4+8 a_1^2 a_2^2)}{64 (J/\hat{J})^6} + \ldots,
\end{multline}
so \Eqs{eq:G12} lead to
\begin{gather}
G_1 = \frac{1}{2 j^2} + \frac{3 z}{2 j^4} + \frac{5(1 + 16z^2)}{16 j^6} + \ldots, \label{eq:G1as}\\
G_2 = \frac{z}{j^2} + \frac{3}{8j^4} + \frac{5 z}{2j^6} + \ldots \label{eq:G2as}
\end{gather}

Note also that, to the extent that $z$ can be neglected, \Eq{eq:rj} can be cast in the following transparent form,
\begin{gather}
\mc{E} \approx \frac{(P')^2}{2m} + \frac{e^2k^2\phi_1^2}{4m(k P'/m)^2}.
\end{gather}
Here $P' \doteq - kJ$, which serves as the canonical OC momentum in $\mc{K}'$. [This is seen if one writes \Eq{eq:P} in $\mc{K}'$, where the frequency is zero.] Hence, the second term on the right-hand side is recognized as the ponderomotive potential \eq{eq:HPhi} produced by a zero-frequency wave in  $\mc{K}'$, and the above formula for $\mc{E}$ is recognized as an approximate formula for the canonical OC energy, \Eq{eq:Hpond}.

\begin{figure*}[t]
\centering
\includegraphics[width=1\textwidth]{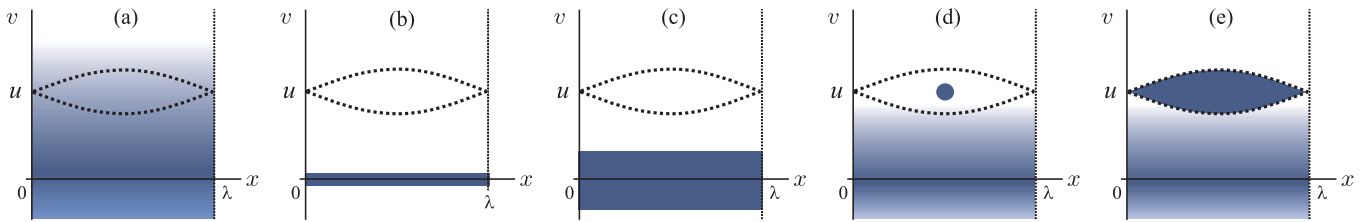}
\caption{Schematics of the specific phase space distributions that are discussed in the present paper: (a) arbitrary smooth distribution; (b) cold limit; (c) finite-temperature waterbag (flat) distribution; (d) deeply trapped particles; (e) flat distribution of trapped particles. In the cases (d) and (e), details of the passing particle distribution are inessential as long as the beam nonlinearity (\Sec{sec:trapped}) dominates. The dashed line is the boundary (separatrix) separating passing trajectories (outside) from trapped trajectories (inside). Also, $x$ is the coordinate, $\lambda \doteq 2\pi/k$, and $v \doteq w + u$ is the velocity in the laboratory frame, $\mc{K}$.}
\label{fig:distr}
\end{figure*}

%%%%%%%%%%%%%%%%%%%%%%%%%%%%%%%%%%%%%%% 
\section{Smooth distributions}
\label{sec:smooth}

%--------------------------------------
\subsection{Basic equations}

First, let us suppose that $F$ is smooth compared to $G$, \ie has a characteristic scale much larger than $J_*$ (\Fig{fig:distr}a), at least near the resonance. This allows approximating $\mc{G}_{1,2}$ as shown in \App{app:smooth}. Specifically, using \Eq{eq:mcGfin} with coefficients $c_q$ inferred by comparing \Eq{eq:Gasymgen} with \Eqs{eq:G1as} and \eq{eq:G2as}, one obtains
\begin{multline}\label{eq:G1apr}
\mc{G}_1
\approx \frac{a_p}{2} \bigg\{[1-\epsilon(\omega, k)] 
 - \omega^2\,\frac{\pd^2\epsilon(\omega, k)}{\pd \omega^2}\,\frac{z a}{2}
\\ -\omega^4\,\frac{\pd^4\epsilon (\omega, k)}{\pd \omega^4}\,\frac{a^2}{192}
+ 4\eta_1\sqrt{a}\, u^3 f_0''(u)\,\frac{\omega_p^2}{\omega^2} \bigg\},
\end{multline}
\begin{multline}\label{eq:G2apr}
\mc{G}_2
\approx a_p\bigg\{[1-\epsilon(\omega, k)] z
- \omega^2\,\frac{\pd^2\epsilon(\omega, k)}{\pd \omega^2}\,\frac{a}{16}
\\ + 2 \eta_2\sqrt{a}\, u^3 f_0''(u)\,\frac{\omega_p^2}{\omega^2}\bigg\},
\end{multline}
where we kept only the terms that will be relevant for our discussion (the applicability conditions will be discussed below; also see \App{app:smooth}) and introduced
\begin{gather}
\eta_1 \doteq \int_0^\infty [\Psi_1(j, z)j - 1/2]\, dj,\\
\eta_2 \doteq \int_0^\infty [\Psi_2(j, z)j - z]\, dj,
\end{gather}
where $\Psi_{1,2}(j, z) \doteq -\int_0^j G_{1,2} (r(\tilde{\jmath}), z)\,d\tilde{\jmath}$. Hence, the NDR [\Eqs{eq:ndr2}] can be cast as follows:
\begin{multline}
0 = \epsilon(\omega, k) 
+ \omega^2\, \frac{\partial^2\epsilon(\omega, k)}{\partial \omega^2}\,\frac{z a}{2}
\\+ \omega^4\,\frac{\partial^4\epsilon(\omega, k)}{\partial \omega^4}\,\frac{a^2}{192}
- 4\eta_1\sqrt{a}\,u^3f_0''(u)\,\frac{\omega_p^2}{\omega^2},
\end{multline}
\begin{gather}
0 = 3z + \omega^2\,\frac{\partial^2\epsilon(\omega, k)}{\partial \omega^2}\,\frac{a}{16}-2\eta_2\sqrt{a}\,u^3 f_0''(u)\,\frac{\omega_p^2}{\omega^2}.
\end{gather}

The linear frequency is then found as a solution of 
\begin{gather}\label{eq:linom}
\epsilon(\omega_0, k) = 0.
\end{gather}
Also one finds that $z = o(1)$, namely,
\begin{gather}\label{eq:zmain}
z \approx - \omega_0^2\,\frac{\partial^2\epsilon(\omega_0, k)}{\partial \omega_0^2}\,\frac{a_0}{48}
+\frac{2}{3}\,\eta_2\sqrt{a_0}\,f_0''(u_0)\,\frac{\omega_0\omega_p^2}{k^3}
\end{gather}
(where $u_0 \doteq \omega_0/k$), and the nonlinear nonlinear frequency shift, $\delta \omega \doteq \omega - \omega_0$, is given by
\begin{multline}\label{eq:dommain}
\delta \omega \approx \bigg[\frac{\partial \epsilon(\omega_0, k)}{\partial \omega_0}\bigg]^{-1}\bigg[
- \omega_0^2\, \frac{\partial^2\epsilon(\omega_0, k)}{\partial \omega_0^2}\,\frac{z a_0}{2}
\\ - \omega_0^4\,\frac{\partial^4\epsilon(\omega_0, k)}{\partial \omega_0^4}\,\frac{a_0^2}{192}
 + 4\eta_1\sqrt{a_0}\,f_0''(u_0)\,\frac{\omega_0\omega_p^2}{k^3}\bigg].
\end{multline}
The coefficients $\eta_{1,2}$ can be evaluated at $z = 0$:
\begin{gather}\label{eq:eta1main}
\eta_1 \approx \int_0^\infty [\Psi_1(j, 0)j - 1/2]\, dj \approx - 0.27,
\end{gather}
which is taken from \Ref{my:actii}, and
\begin{multline}\label{eq:eta2main}
\eta_2 
\approx \int_0^\infty \Psi_2(j, 0)j\, dj
= \frac{1}{2}\int_0^\infty G_2(r(j, 0), 0)j^2\, dj \\
= \frac{1}{2}\int_0^\infty G_2(r, 0)j^2(r, 0)\,j_r(r, 0)\,dr \approx 0.11,
\end{multline}
which is calculated numerically. Below, we demonstrate how these equations are applied to specific distributions and show how our results correspond to those that were reported in the literature previously.

%--------------------------------------
\subsection{Fluid limit}
\label{sec:fluid}

Let us start with the fluid limit, when $k \lambda_D$ is relatively small (here $\lambda_D \doteq v_T/\omega_p$ is the Debye length, and $v_T$ is the thermal speed), while $a$ is relatively large (although $a \ll 1$ is still assumed), such that one can drop the half-integer powers of $a_0$ in the above formulas. In particular, \Eq{eq:zmain} then yields
\begin{gather}\label{eq:zfluid}
z = - \omega_0^2\,\frac{\partial^2\epsilon(\omega_0, k)}{\partial \omega_0^2}\,\frac{a_0}{48}.
\end{gather}
By substituting this into \Eq{eq:dommain} and again ignoring the term $a_0^{1/2}$, we obtain
\begin{multline}\label{eq:wfluid}
\delta \omega \approx \frac{a_0^2}{96}\,
\bigg[\frac{\partial \epsilon(\omega_0, k)}{\partial \omega_0}\bigg]^{-1}\\
\times \bigg\{
\bigg[\omega_0^2\, \frac{\partial^2\epsilon(\omega_0, k)}{\partial \omega_0^2}\bigg]^2 -
\frac{\omega_0^4}{2}\,\frac{\partial^4\epsilon(\omega_0, k)}{\partial \omega_0^4}
\bigg\}.
\end{multline}

The effect $\delta \omega \propto a^2$ is called a fluid nonlinearity. Note, however, that trapped particles also contribute to this nonlinearity in the general case, contrary to a common misconception. This is because, unless $f_0$ is vanishingly small near the resonance, removing trapped particles would leave a phase space hole, rendering $f_0$ abrupt and the above results inapplicable. (The corresponding modification of the NDR can be quite strong and is discussed in \Sec{sec:trapped}.) What makes the plasma act as a fluid then is not necessarily the absence of trapped particles but rather the smoothness of $f_0$ near the resonance.
 
To our knowledge, \Eq{eq:wfluid} is the first generalization of fluid nonlinearities to plasmas with arbitrary $\epsilon$. Below we consider several special cases for illustration.

%--------------------------------------
\subsubsection{Cold limit}
\label{sec:cold}

In the limit of cold plasma that is initially at rest (\Fig{fig:distr}b), one has
\begin{gather}\label{eq:epscold}
\epsilon(\omega, k) = 1 - \omega_p^2/\omega^2.
\end{gather}
Then \Eq{eq:linom} gives $\omega_0 = \omega_p$, and \Eqs{eq:zfluid} and \eq{eq:wfluid} give
\begin{gather}\label{eq:fluidzw}
z = a_0/8, \quad \delta \omega = \omega_p a_0^2/2.
\end{gather}

This result is in agreement with \Ref{ref:dewar72e}, but it may seem to be at variance with the widely known theorem \cite{foot:sagdeev} that a Langmuir wave in cold plasma exhibits no nonlinear frequency shift whatsoever (in fact, even at large $a_0$). The discrepancy is due to the fact that the mentioned theorem applies to the plasma rest frame only. Our calculation, in contrast, is done in the laboratory frame, where the excitation of a plasma wave is accompanied by plasma acceleration. That results in the nonlinear nonlinear Doppler shift discussed in \Sec{sec:doppler}. One can easily check, by substituting \Eq{eq:epscold} into \Eq{eq:ndop}, that $\delta \omega$ described by \Eq{eq:fluidzw} entirely consists of this shift. Hence, recalculating \Eq{eq:fluidzw} to the plasma rest frame predicts zero frequency shift, as anticipated. 

%--------------------------------------
\subsubsection{Waterbag distribution}

Now suppose the waterbag distribution discussed in \Ref{ref:winjum07}. Specifically, assume that the initial velocities of particles are distributed homogeneously within the interval $(-\bar{v}, \bar{v})$, where $\bar{v}$ is some constant (\Fig{fig:distr}c). The corresponding dielectric function is calculated by directly taking the integral in \Eq{eq:eps0} \cite{foot:waterbag} and is given by
\begin{gather}
\epsilon(\omega, k) = 1 - \frac{\omega_p^2}{\omega^2 - k^2 \bar{v}^2}.
\end{gather}
Then \Eq{eq:linom} yields $\omega_0^2 = \omega_p^2 + k^2 \bar{v}^2$, and \Eqs{eq:zfluid} and \eq{eq:wfluid} lead to
\begin{gather}
z \approx \frac{a_0(3 + \bar{\alpha})}{24(1 - \bar{\alpha})^2},\\
\delta \omega \approx \omega_0 a_0^2\,\frac{(6 + 9 \bar{\alpha} + \bar{\alpha}^2)}{12(1 - \bar{\alpha})^3},\label{eq:wb}
\end{gather}
where $\bar{\alpha} \doteq (k\bar{v}/\omega_0)^2$. At $\bar{\alpha} \to 0$, \Eq{eq:wb} reproduces the result that we presented in \Sec{sec:cold} for cold plasma.

One may notice that \Eq{eq:wb} is at variance with the result obtained in \Ref{ref:winjum07}. This is due to the fact that, in \Ref{ref:winjum07}, the NDR is calculated not in the laboratory frame but rather in the frame where the average velocity is zero; hence the nonlinear Doppler shift is not included. For the distribution in question, \Eq{eq:ndop} gives
\begin{gather}\notag
k \ffavr{\Delta V} \approx \frac{\omega_0 a^2}{2(1-\bar{\alpha})^2}.
\end{gather}
Subtracting this from \Eq{eq:wb} leads to $\delta\omega = \omega_0 a^2 \bar{\alpha} (15 + \bar{\alpha})/[12(1 - \bar{\alpha})^3]$, which reproduces the result of \Ref{ref:winjum07}.

%--------------------------------------
\subsubsection{Kappa distribution}

\begin{figure*}
\centering
\includegraphics[width=.49\textwidth]{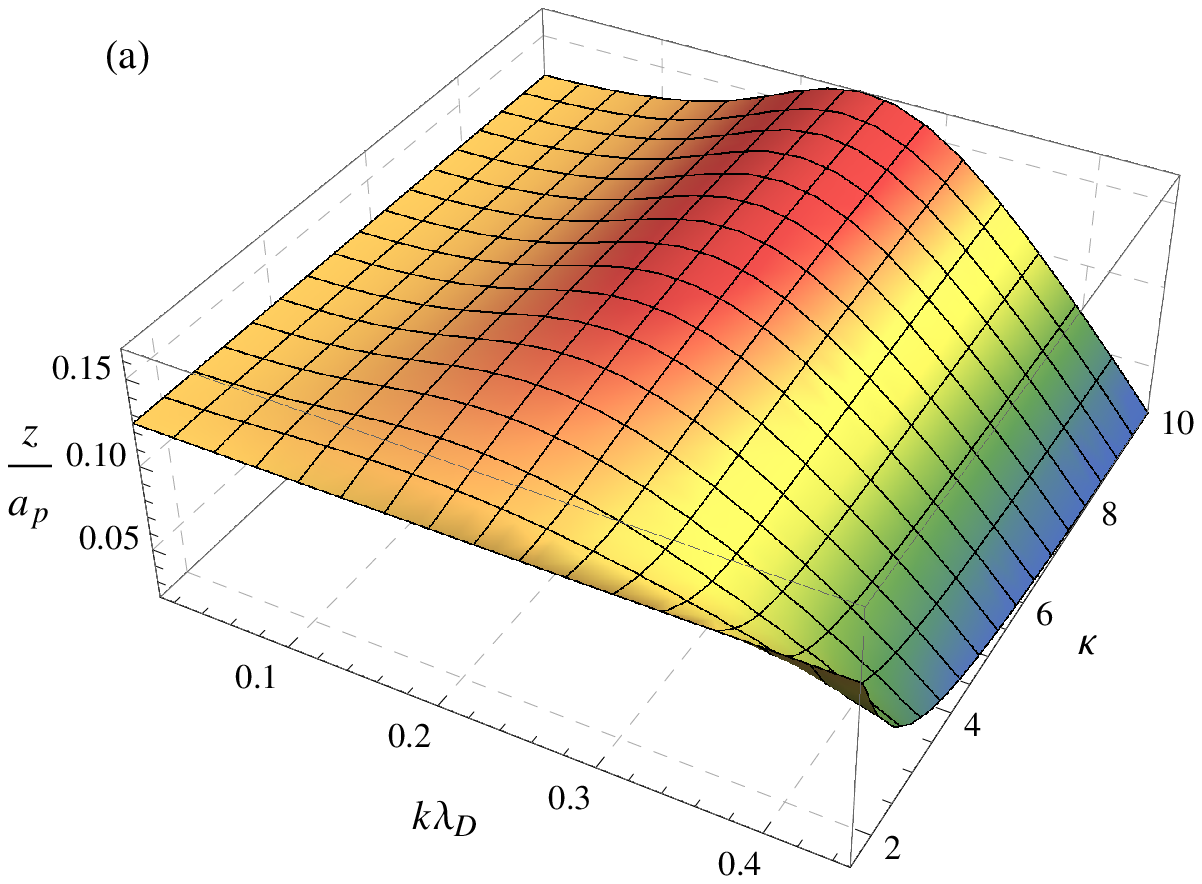}
\hfill
\includegraphics[width=.49\textwidth]{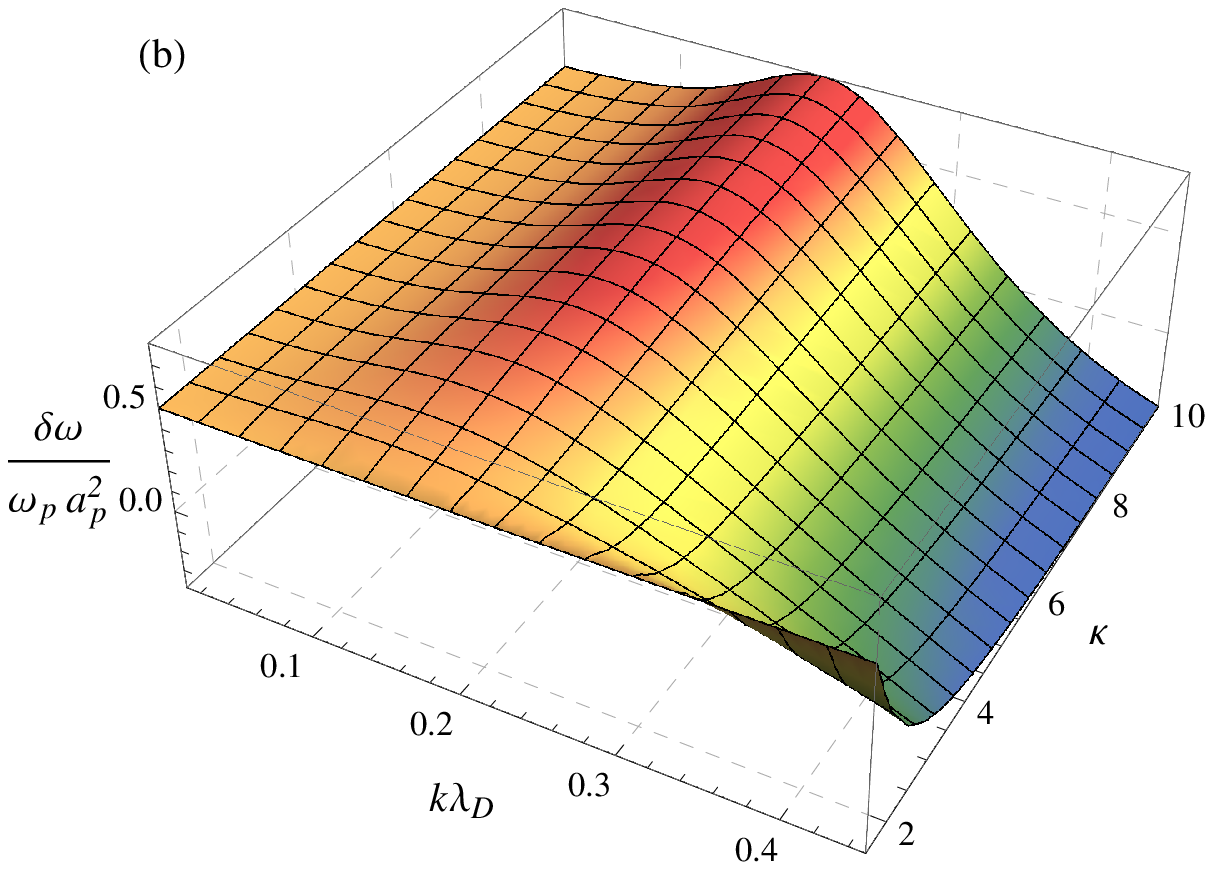}
\caption{(Color online) The effect of the fluid nonlinearity [\Eqs{eq:zfluid} and \eq{eq:wfluid}] on the dispersion of a nonlinear Langmuir wave in electron plasma with the kappa distribution, \Eq{eq:kappadistr}: (a) normalized amplitude of the second harmonic, $z \doteq \phi_2/(2\phi_1)$, in units $a_p$; (b) nonlinear frequency shift, $\delta\omega$, in units $\omega_p a_p^2$. The horizontal axes are $k\lambda_D$ and the distribution parameter $\kappa$. At $\kappa \gg 1$, the kappa distribution is approximately Maxwellian.}
\label{fig:kappa}
\end{figure*}

We also calculated the effect of the fluid nonlinearity on the dispersion of a nonlinear Langmuir wave \cite{foot:eaw} in electron plasma with the kappa distribution \cite{ref:livadiotis13},
\begin{gather}\label{eq:kappadistr}
f_0(V_0) = \frac{\Gamma(\kappa + 1)/\Gamma(\kappa - 1/2)}{v_T\kappa\sqrt{\pi(2\kappa - 3)}} \left[1 + \frac{V_0^2}{(2\kappa - 3)v_T^2}\right]^{-\kappa},
\end{gather}
where $\kappa > 3/2$ is a constant dimensionless parameter, $v_T^2 = \int^\infty_{-\infty} v^2 f_0(v)\,dv$ is the thermal speed squared, and $\Gamma$ is the gamma function. [The distribution \eq{eq:kappadistr} approaches the Maxwellian distribution at $\kappa \gg 1$, but has more pronounced tails at smaller $\kappa$.] The results are shown in \Fig{fig:kappa}. Notice that $z$ is not monotonic in $k$, and $\delta \omega$ can have either sign. This is in contrast to \Ref{ref:winjum07}, where the fluid frequency shift was reported as strictly positive. 

%--------------------------------------
\subsection{Kinetic limit}
\label{sec:kinetic}

Now let us consider the opposite limit, when the amplitude is relatively small, whereas $k \lambda_D$ is substantial. In this case,
\begin{gather}
z = \frac{2}{3}\,\eta_2\sqrt{a_0}\,f_0''(u_0)\,\frac{\omega_0\omega_p^2}{k^3}, \label{eq:zkin}\\
\delta \omega \approx 4\eta_1\sqrt{a_0}\,f_0''(u_0)\,\frac{\omega_0\omega_p^2}{k^3}
\bigg[\frac{\partial \epsilon(\omega_0, k)}{\partial\omega_0}\bigg]^{-1}. \label{eq:omkin}
\end{gather}
The effect $\delta \omega \propto \sqrt{a_0}$ is called a kinetic nonlinearity (associated with a smooth distributions; for abrupt distributions, see \Sec{sec:trapped}). In particular, \Eq{eq:zkin} shows that $\phi_2 \propto \phi_1^{3/2}$. This is in qualitative agreement with the (not self-consistent) estimate in \Ref{ref:rose01}, but notice that our numerical coefficient is different. As regarding \Eq{eq:omkin}, it is in precise agreement (modulo typos) with \Refs{my:actii, my:bgk} and also with the results reported in \Ref{ref:dewar72b} for the ``adiabatic excitation''. 

\begin{figure*}
\centering
\includegraphics[width=.49\textwidth]{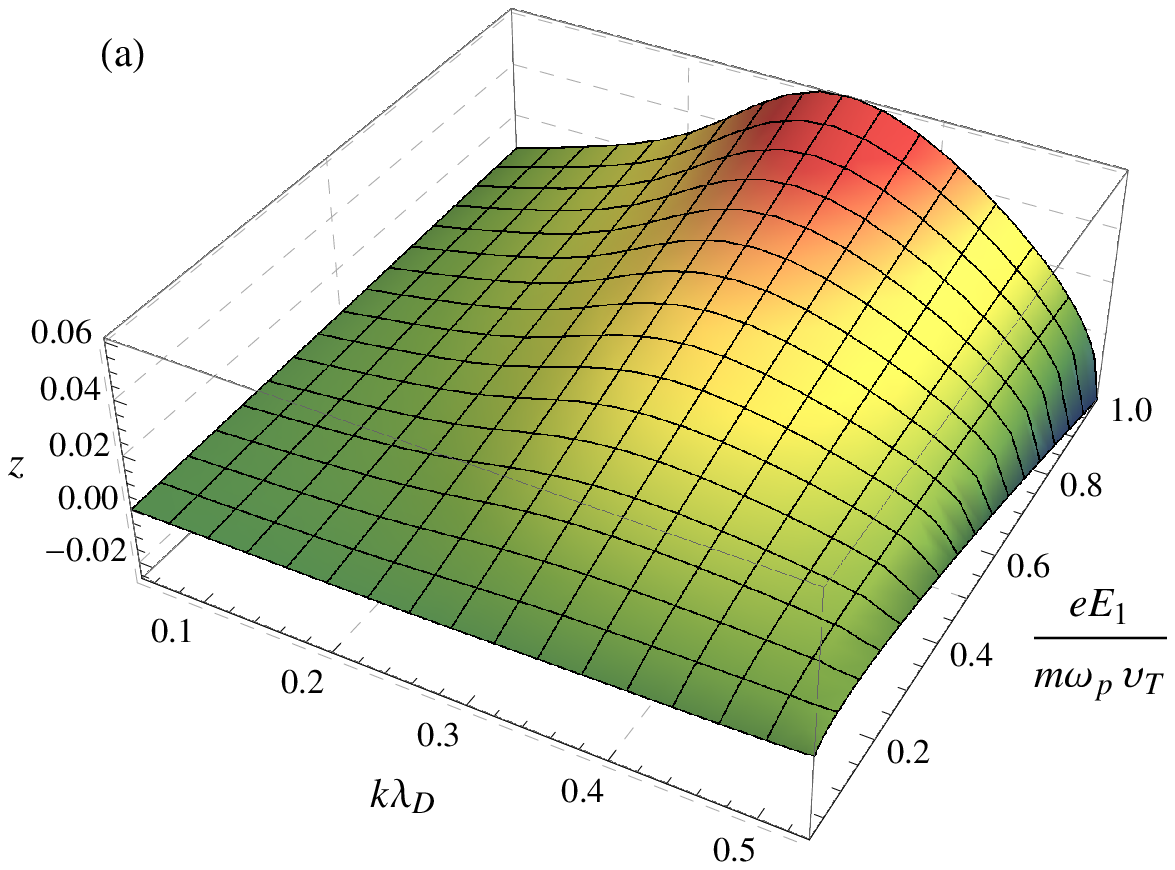}
\hfill
\includegraphics[width=.49\textwidth]{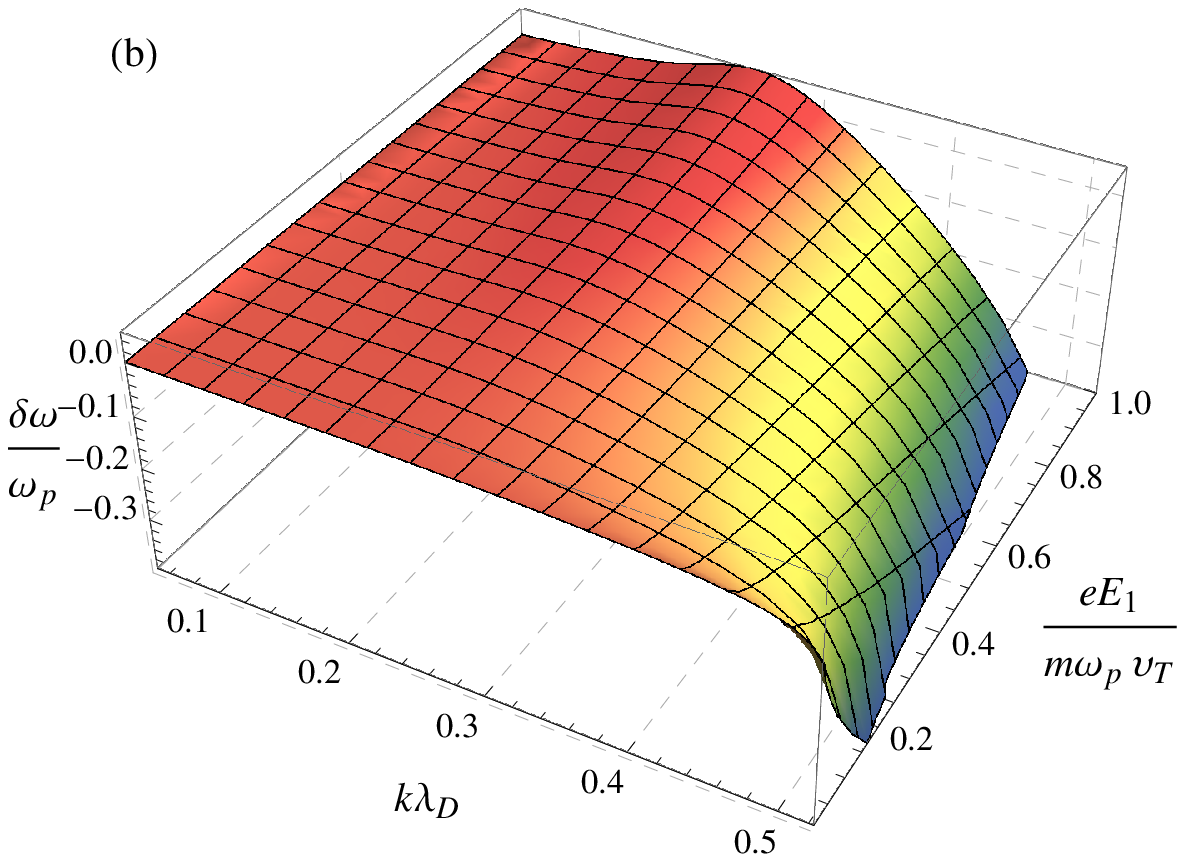}
\caption{(Color online) The combined effect of the fluid and kinetic nonlinearities on the dispersion of a nonlinear Langmuir wave in electron plasma with the Maxwellian distribution, \Eq{eq:maxwellf}: (a) normalized amplitude of the second harmonic, $z \doteq \phi_2/(2\phi_1)$; (b) nonlinear frequency shift, $\delta\omega$, in units $\omega_p$. The horizontal axes are $k\lambda_D$ and $eE_1/(m\omega_p v_T) = v_{\rm osc}/v_T$, where $E_1 \doteq k\phi_1$ is the amplitude of the first harmonic of the wave electric field, and $v_{\rm osc} \doteq eE_1/(m\omega_p)$ is the characteristic amplitude of the electron velocity oscillations. The results are obtained by application of the approximate formulas \eq{eq:zmain} and \eq{eq:dommain}.}
\label{fig:maxwell1}
\end{figure*}

%--------------------------------------
\subsection{Fluid and kinetic nonlinearities combined}
\label{sec:combined}

We also calculated the combined effect of the fluid and kinetic nonlinearities for a nonlinear Langmuir wave \cite{foot:eaw} in electron plasma with the Maxwellian distribution,
\begin{gather}\label{eq:maxwellf}
f_0(V_0) = \frac{1}{v_T\sqrt{2\pi}} \exp\left(-\frac{V_0^2}{2v_T^2}\right).
\end{gather}
In this case,
\begin{gather}
\epsilon(\omega, k) = 1 - \frac{1}{2k^2 \lambda^2}\,Z'\left(\frac{\omega}{kv_T\sqrt{2}}\right),
\end{gather}
where $Z(\zeta) = - 2S(\zeta)$, and $S$ is the Dawson function, $S(\zeta) \doteq \exp(-\zeta^2) \int_0^\zeta \exp (y^2)\, dy$ \cite[Chap.~8]{book:stix}. 

\begin{figure*}
\centering
\includegraphics[width=.9\textwidth]{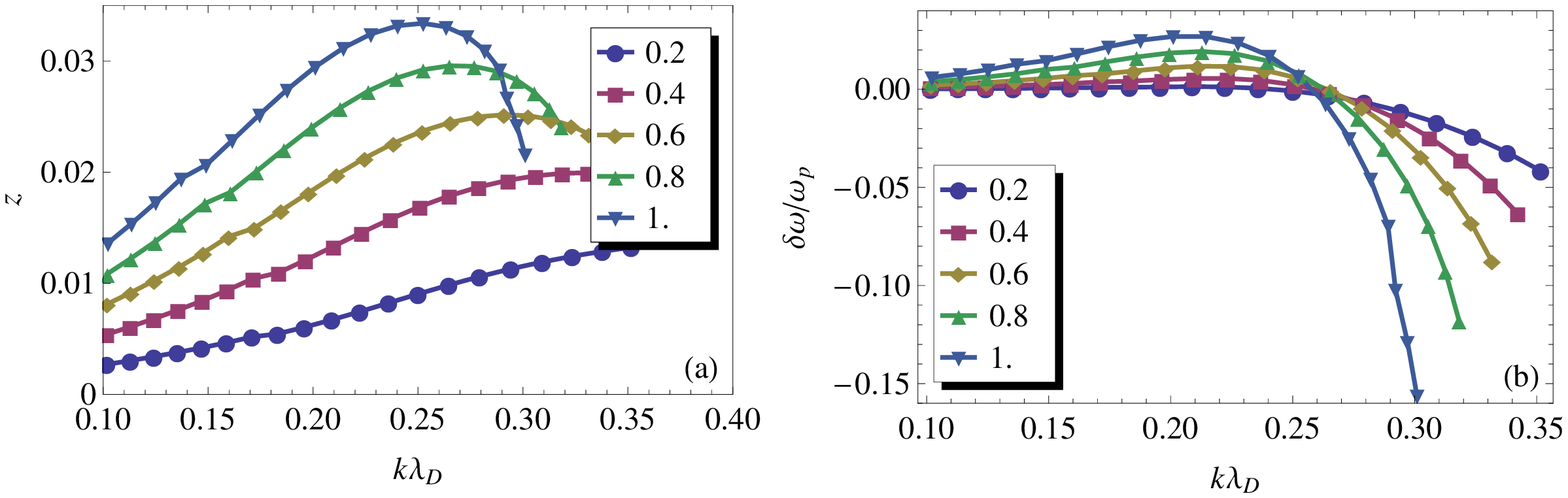}
\caption{(Color online) Same as in \Fig{fig:maxwell1} but using the more precise NDR, \Eq{eq:ndr2}, and only for selected values of $eE/(m \omega_p v_T)$ that are specified in the legend.}
\label{fig:maxwellex}
\end{figure*}

Two sets of results are presented. Figure \ref{fig:maxwell1} shows the results obtained by application of the asymptotic formulas \eq{eq:zmain} and \eq{eq:dommain}. Figure \ref{fig:maxwellex} shows the result of the direct numerical solution of the more precise NDR, \Eq{eq:ndr2}. It is seen that the asymptotic formulas provide a reasonably accurate approximation to the exact solution, even though the applicability conditions of the asymptotic theory, \eq{eq:appappl}, are satisfied only marginally. In particular, the following trends are seen. At small $k\lambda_D$, the fluid  nonlinearity dominates, leading to \Eqs{eq:fluidzw}. (The fact that, in the figure, $\delta\omega$ remains finite at vanishing $k\lambda_D \propto v_T$ is due to the choice of the units in which the amplitude is measured.) At larger $k\lambda_D$, the kinetic nonlinearity becomes dominating. That said, comparison with \Fig{fig:kappa} shows that the presence of the kinetic nonlinearity does not affect the picture qualitatively.

%%%%%%%%%%%%%%%%%%%%%%%%%%%%%%%%%%%%%%% 
\section{Beam distributions}
\label{sec:trapped}

In addition to fluid and kinetic nonlinearities discussed above, waves with trapped particles can also exhibit nonlinearities of a third, beam type. Those emerge when, \textit{on top} of a distribution $F_0$ that is smooth or negligibly small near the resonance, an additional distribution $F_\msf{b}$ of a trapped beam is superimposed that has abrupt boundaries within or at the edge of the trapping island. Such distributions, with both positive $F_\msf{b}$ (clumps) and negative $F_\msf{b}$ (holes), are known in various contexts \cite{ref:goldman71, ref:krasovsky94, ref:krasovsky92, ref:krasovskii95, ref:friedland06, ref:krasovsky07, ref:breizman10, my:actiii, my:trcomp, my:nmi}. Below, we consider some of them in detail.

%--------------------------------------
\subsection{Basic equations}

The presence of $F_\msf{b}$ results in the appearance of additional terms,
\begin{gather}\label{eq:tildeG}
\tilde{\mc{G}}_{1,2} \doteq \int G_{1,2} F_\msf{b}(J)\,dJ,
\end{gather}
in \Eqs{eq:G1apr} and \eq{eq:G2apr}. For simplicity, we will suppose that $F_\msf{b}$ is large enough, such that its nonlinear effect on the NDR dominates over the nonlinear effect produced by $F_0$. Then,
\begin{gather}
\mc{G}_1 \approx (a_p/2)[1-\epsilon(\omega, k)] + \tilde{\mc{G}}_1,\label{eq:Gcl}\\
\mc{G}_2 \approx a_p z[1-\epsilon(\omega, k)] + \tilde{\mc{G}}_2,
\end{gather}
so \Eqs{eq:ndr2} yield the NDR in the form
\begin{gather}\label{eq:aux47}
z \approx \frac{\tilde{\mc{G}}_2}{3a_p}, \quad 
\delta \omega \approx \frac{2\tilde{\mc{G}}_1}{a_p} \left[\frac{\pd\epsilon(\omega_0, k)}{\pd\omega_0}\right]^{-1}.
\end{gather}
Notice that the wave is close to linear in this case not when $a_p$ is small, like before, but rather when $\beta \doteq \bar{n}_\msf{b}/(\bar{n}a_p)$ is small, where $\bar{n}_\msf{b}$ is the beam density. (In other words, $a_p$ must be sufficiently \textit{large}.) The NDRs for specific distributions are derived as follows.

%--------------------------------------
\subsection{Deeply trapped particles}

Suppose that trapped particles reside at the very bottom of the wave troughs (\Fig{fig:distr}d); \ie
\begin{gather}\label{eq:deepF}
F_\msf{b}(J) = (\bar{n}_\msf{b}/\bar{n})\,\delta(J).
\end{gather}
[Strictly speaking, the delta function here must be understood as the limit of $\delta(J-J_c)$ at $J_c \to 0$.] Hence, all trapped particles have $r = z$, so \Eqs{eq:glim0}, \eq{eq:tildeG}, and \eq{eq:deepF} yield $\tilde{\mc{G}}_{1,2} = \mp \bar{n}_\msf{b}/\bar{n}$. Then, \Eqs{eq:aux47} give
\begin{gather}
\label{eq:aux48}
z \approx \frac{\beta}{3},
\quad
\frac{\delta \omega}{\omega_p} \approx - \frac{2\beta}{\omega_p}\left[\frac{\pd\epsilon(\omega_0, k)}{\pd\omega_0}\right]^{-1}.
\end{gather}
The obtained expression for $\delta \omega$ agrees with the well known result \cite{ref:goldman71, ref:krasovsky94}. It is also easy to see that the value of $z$ is in agreement with what one can infer from the Vlasov-Poisson system directly \cite{foot:calc}.

%--------------------------------------
\subsection{Homogeneous clumps and holes}
\label{sec:clump}

%--------------------------------------
\subsubsection{Basic equations}

Let us also consider the case when $F_\msf{b}$ is flat across the whole trapping area (\Fig{fig:distr}e). In this case, $F_\msf{b}(J < J_*) = \hat{F}$, where $\hat{F}$ is some constant. Then, 
\begin{multline}\label{eq:mcG2}
\tilde{\mc{G}}_{1,2} = \hat{J}\sqrt{a}\hat{F} \int_z^{1+z} G_{1,2}(r, z)\, j_r(r, z)\,dr \\
=  a_p \mc{N} \gamma_{1,2}(z) \approx a_p \mc{N} \gamma_{1,2}(0),
\end{multline}
where we introduced the dimensionless quantities
\begin{gather}
\mc{N} \doteq \frac{m \omega_p}{k}\,\frac{\hat{F}}{\sqrt{a_p}} \sim \beta,\label{eq:mcN}\\
\gamma_{1,2}(z) \doteq \int_z^{1 + z} G_{1,2}(r, z)\, j_r(r, z)\,dr.
\end{gather}
Hence, one obtains
\begin{gather}
z \approx \frac{1}{3}\,\gamma_2(0)\mc{N}, \label{eq:clumpz}\\
\delta \omega \approx 2\gamma_1(0)\mc{N}\left[\frac{\pd\epsilon(\omega_0, k)}{\pd\omega_0}\right]^{-1}.\label{eq:clumpw}
\end{gather}
We find numerically that $\gamma_2(0) \approx -0.085$, and $\gamma_1(0) = -4/(3\pi)$ was reported already in \Ref{my:actii}.

%--------------------------------------
\subsubsection{Discussion}

Since $\tilde{\mc{G}}_1$ given by \Eq{eq:mcG2} is independent of $\omega$, one can also derive a more precise analytic expression for $\omega$ from \Eq{eq:Gcl} when $\epsilon$ is a simple enough function. For instance, for cold plasma [\Eq{eq:epscold}], one gets
\begin{gather}\label{eq:clumpwcold}
\frac{\omega}{\omega_p}  = \left(1 + \frac{8\mc{N}}{3\pi}\right)^{\!-1/2}\!\!\! = 1 -\frac{4\mc{N}}{3\pi}+\frac{8\mc{N}^2}{3\pi^2} + O(\mc{N}^3).
\end{gather}
Notably, phase space clumps ($\mc{N} > 0$) in this case correspond to $\delta \omega < 0$, whereas phase space holes ($\mc{N} < 0$) correspond to $\delta \omega > 0$.

Let us compare results predicted by \Eq{eq:clumpwcold} for $\omega$, as well as those by \Eq{eq:clumpz} for $z$, with the exact kinetic solution, which happens to exist in this special case. [The word ``exact'' here refers to the description of trapped particles, whereas the bulk plasma is still modeled with a linear dielectric function \eq{eq:epscold}.] Assuming the notation $\alpha \doteq (\pi/2)(\omega_p/\omega)$, the exact solution is given by \cite{foot:brz}
\begin{gather}
z = \frac{\pi^2 - 4\alpha^2}{32\pi^2 - 8 \alpha^2},\label{eq:brz}\\
\mc{N}^{-2} = \frac{2\tan\alpha}{3\alpha}\,\frac{\pi^4}{\pi^4 - 5\pi^2\alpha^2 + 4\alpha^4},\label{eq:brw}
\end{gather}
or, more explicitly for small $\mc{N}$,
\begin{gather}
z = -\frac{4 \mc{N}}{45 \pi} + O(\mc{N}^2),\\
\frac{\omega}{\omega_p} = 1 - \frac{4\mc{N}}{3\pi} + \frac{68\mc{N}^2}{27\pi^2} + O(\mc{N}^3).
\end{gather}

\begin{figure}
\centering
\includegraphics[width=.48\textwidth]{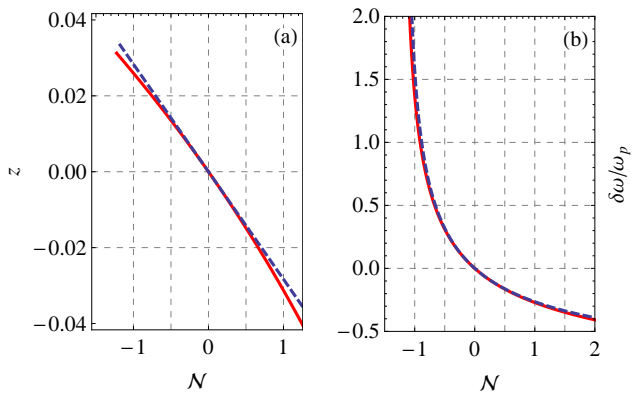}
\caption{(Color online) The effect of the beam nonlinearity on the dispersion of a nonlinear Langmuir wave in cold electron plasma for the case when the trapping islands contain homogeneous clumps ($\mc{N} > 0$, as in \Fig{fig:distr}e) or holes ($\mc{N} < 0$): (a) normalized amplitude of the second harmonic, $z \doteq \phi_2/(2\phi_1)$; (b) nonlinear frequency shift, $\delta\omega$, in units $\omega_p$. The coordinate $\mc{N}$, which is given by \Eq{eq:mcN}, represents a dimensionless measure of the trapped-particle phase space density. The red solid curves correspond to the exact solutions, \Eqs{eq:brz} and \eq{eq:brw}. The blue dashed curves correspond to our asymptotic solutions, \Eqs{eq:clumpz} and \eq{eq:clumpwcold}. There is no solution for $z$ and $\omega$ at $\mc{N} < \mc{N}_c$. The exact theory predicts $\mc{N}_c \approx -1.22$, and the asymptotic theory predicts $\mc{N}_c \approx -1.18$.}
\label{fig:clump}
\end{figure}

As seen in \Fig{fig:clump}, the results of our asymptotic theory are virtually indistinguishable from the predictions of \Eqs{eq:brz} and \eq{eq:brw} at $|\mc{N}| \lesssim 1$ and even beyond. The asymptotic theory also captures the fact that there is no solution for $z$ and $\omega$ below some threshold, $\mc{N} < \mc{N}_c$; \ie deep enough hole modes cannot stable \cite{foot:eremin}. Our \Eq{eq:clumpwcold} predicts $\mc{N}_c = - 3\pi/8 \approx -1.18$. This differs by less than 4\% from the exact-theory prediction, $\mc{N}_c = - \sqrt{3/2} \approx - 1.22$, which is obtained from \Eq{eq:brw} in the limit $\alpha \to 0$, \ie at $\omega/\omega_p \to \infty$.

One conclusion from this is that the long-range sweeping of energetic-particle modes, which is described in \Ref{ref:breizman10} by means of the aforementioned exact solution, can be described with fidelity just using a simple asymptotic NDR, \Eq{eq:clumpwcold}. [In fact, \Eq{eq:clumpwcold} was proposed already in \Ref{ref:krasovsky07}.] Using the theory presented here, asymptotic NDRs can be derived also when exact analytic solutions do not exist. For example, in addition to beam nonlinearities, one may need to account for fluid nonlinearities when $a^2_p \sim \mc{N}$, and for kinetic nonlinearities when $a_p \sim \mc{N}^2$ and $k \lambda_D \sim 1$. That is done simply by retaining \textit{all} the terms in the expressions for $\mc{G}_{1,2}$, including not only $\tilde{\mc{G}}_{1,2}$ but also the other terms that appear in \Eqs{eq:G1apr} and \eq{eq:G2apr}. To the leading order, the nonlinear effects stemming from fluid, kinetic, and beam effects then enter the NDR additively.

%%%%%%%%%%%%%%%%%%%%%%%%%%%%%%%%%%%%%%% 
\section{Conclusions}
\label{sec:conc}

In summary, we derived a transparent asymptotic expression for the nonlinear frequency shift of intense Langmuir waves in general collisionless plasma. Our result describes kinetic and fluid effects simultaneously. This is achieved by application of a variational approach similar to the approach used in \Refs{my:bgk, my:itervar, my:acti, my:actii}. In contrast to the previous calculations, however, the present one takes into account corrections to the NDR caused by the wave second harmonic. It is shown for the first time that the fluid nonlinearity, which is determined by the second harmonic, can be cast in terms of the plasma dielectric function [\Eq{eq:wfluid}] and, contrary to the common presumption, can have either sign. Also, the kinetic nonlinearity that we derive accounts for both smooth distributions [\Eq{eq:omkin}] and trapped-particle beams (\Sec{sec:trapped}). Our formulation is benchmarked against the many previously known NDRs and is shown to reproduce them as special cases of a single unifying theory. For example, we calculate the NDRs for Langmuir waves in plasmas with cold, waterbag, kappa, and Maxwellian distributions. These specific results and our general method are applicable, \eg to waves produced at intense laser-plasma interactions and to energetic-particle modes in tokamaks. Our asymptotic formulation, in fact, may be advantageous over exact kinetic solutions for such waves. While offering a reasonable precision, it leads to simpler (and thus more elucidating) results, does not involve solving any differential equations, and can be extended straightforwardly to other nonlinear plasma waves. 

One of the authors (IYD) thanks D. B\'enisti for valuable discussions. The work was supported by the NNSA SSAA Program through DOE Research Grant No. DE274-FG52-08NA28553 and by the DOE contract DE-AC02-09CH11466.

%%%%%%%%%%%%%%%%%%%%%%%%%%%%%%%%%%%%%%% 
\appendix

%%%%%%%%%%%%%%%%%%%%%%%%%%%%%%%%%%%%%%% 
\section{Low-frequency field}
\label{app:E}

Here, we discuss some issues regarding calculations of the LF field, $\bar{E} \doteq - \pd_x \bar{\phi}$, in electrostatic plasma approximation (EPA). A common way to define such an approximation is to omit, in Ampere's law, the curl of the magnetic field transverse to the current, $\vec{B}_\perp$. In projection on the current axis, $x$, this gives
\begin{gather}\label{eq:amperes}
\pd_t E + 4 \pi j = 0,
\end{gather}
where, unlike in the rest of the paper, $j$ denotes the current density. For instance, this form was used in \Refs{ref:mckinstrie87, ref:lindberg07}. It was concluded then, in particular, that $\pd_t \bar{E} = - 4\pi \favr{j}$, in which case a stationary wave is possible only at zero average current. 

However, strictly speaking, the assumption of one-dimensional dynamics implies that $\vec{B}_\perp$ has a vanishing effect on the particle motion but not a vanishing curl. Those are not equivalent requirements. For instance, consider a cylindrical plasma column with current along the axis of symmetry. On the axis, one can have nonzero component of $\nabla \times \vec{B}_\perp$ along the current even though $\vec{B}_\perp$ itself is locally zero. Off the axis, $\vec{B}_\perp$ is a continuous function of the radius $R$, so its effect is small at small enough $R$, while the curl of $\vec{B}_\perp$ is nonnegligible. (This is a standard assumption, \eg in the plasma probe theory.) Moreover, the effect of $\vec{B}_\perp$ can be made arbitrarily small everywhere, if an additional strong homogeneous magnetic field is imposed parallel to the plasma column. 

This shows that \Eq{eq:amperes} is not always reliable for determining $E$ in the EPA. Gauss's law can be used instead,
\begin{gather}\label{eq:gauss's}
\nabla^2 \phi = - 4 \pi \rho,
\end{gather}
where $\rho$ is the charge density. (Note that, when calculating $\rho$ on envelope scales, one generally needs to account for ponderomotive forces.) But \Eq{eq:gauss's} contains spatial derivatives, so it must be supplemented with boundary conditions. These boundary conditions are determined by external charges at plasma walls (if any), so they are independent of the average current. This means, in particular, that the nonlinear Doppler shift described in \Sec{sec:doppler} may or may not be associated with the build-up of a LF electric field, depending on a particular system. For instance, absent external inductive fields, the average field in case of periodic boundary conditions (\eg in toroidal plasma) will remain zero at all times \cite{foot:period}. 

Note also that \Eq{eq:amperes} generally violates the fundamental energy-momentum conservation in EPA, as discussed in \Ref{foot:qin}. In contrast, the field-theoretical approach developed in the main text, which implies \Eq{eq:gauss's} yet not \Eq{eq:amperes}, predicts that the wave excitation inputs precisely the amount of canonical momentum expected from the linear wave theory (\Sec{sec:doppler}; also see \Ref{my:itervar} for the energy conservation). It \textit{is} possible that, eventually, a LF field is established and redistributes plasma momentum or transfers some of that to the walls; however, as we explain in \Sec{sec:wld}, these effects are unimportant to calculating a local NDR. Given an OC distribution for a specific time of interest, the NDR can be calculated irrespective of the factors that have shaped that given distribution. In this sense, whether $\bar{E}$ is zero or not is irrelevant in the context of this paper.

%%%%%%%%%%%%%%%%%%%%%%%%%%%%%%%%%%%%%%% 
\section{Bulk plasma acceleration by homogeneous wave excitation}
\label{app:hydro}

When a homogeneous electrostatic wave is excited in plasma, the plasma generally changes its average velocity by some $\ffavr{\Delta V}$. One derivation of this effect was presented in \Sec{sec:doppler} using the OC formalism. Here, we show how a hydrodynamic model yields the same result.

Consider the equation for the plasma momentum density, $\mc{P} = mnv$, namely,
\begin{gather}
\pd_t\mc{P} + \pd_x (m n v^2) = n e E - \pd_x \Pi,
\end{gather}
where $n$ is the particle density, $v$ is the flow velocity, and $\Pi$ is the pressure. By averaging over $x$, which we denote with $\favr{\ldots}_x$, one gets
\begin{gather}\label{eq:aux771}
\pd_t \favr{\mc{P}}_x = \favr{n e E}_x \approx \favr{\tilde{n} e \tilde{E}}_x.
\end{gather}
(The tilde in this appendix denotes quantities of the first order in the field amplitude.) Here we used that the gradient of a periodic function averages to zero. We also assumed zero external LF field and periodic boundary conditions, so a LF field remains zero at all $t$ (\App{app:E}). 

Notice that $e \tilde{E}$ that enters \Eq{eq:aux771} can be found from
\begin{gather}\label{eq:Ev}
e \tilde{E} = m \pd_t \tilde{v} + \pd_x \tilde{\Pi}/\bar{n}.
\end{gather}
Assuming $\Pi = \Pi(n)$, we have $\tilde{n}\, \pd_x \tilde{\Pi}/\bar{n} = \text{const} \times \pd_x \tilde{n}^2$,~so
\begin{gather}
\favr{\tilde{n} e \tilde{E}}_x = m \favr{\tilde{n} \pd_t \tilde{v}}_x = m\pd_t \favr{\tilde{n}\tilde{v}}_x - m\favr{\tilde{v} \pd_t \tilde{n}}_x.
\end{gather}
From the continuity equation, we also have $\pd_t \tilde{n} = - \bar{n} \pd_x \tilde{v}$. This leads to
\begin{align}
\favr{\tilde{n} e \tilde{E}}_x 
& = m \pd_t \favr{\tilde{n}\tilde{v}}_x + m \bar{n}\favr{\tilde{v} \pd_x \tilde{v}}_x \notag\\
& = \pd_t \favr{m \tilde{n}\tilde{v}}_x + \favr{\pd_x(m\bar{n}\tilde{v}^2/2)}_x \notag\\
& = \pd_t \favr{m \tilde{n}\tilde{v}}_x.
\end{align}
Thus, $\pd_t\favr{\mc{P}}_x = \pd_t \favr{m \tilde{n} \tilde{v}}_x$, or, after integrating over $t$,
\begin{gather}\label{eq:auxP}
\favr{\Delta \mc{P}}_x = \favr{m \tilde{n} \tilde{v}}_x = m u \favr{\tilde{n}^2}_x/\bar{n}= m \bar{n} \favr{\tilde{v}^2}_x/u,
\end{gather}
where we also substituted the continuity equation, $\tilde{n} \approx \bar{n} \tilde{v}/u$. Then, $\ffavr{\Delta V} \doteq \favr{\Delta \mc{P}}_x/(m\bar{n})$ is given by
\begin{gather}\label{eq:auxV}
\ffavr{\Delta V} = u \favr{(\tilde{n}/\bar{n})^2}_x = \favr{\tilde{v}^2}_x/u.
\end{gather}
Since, in the assumed model, one has $\epsilon(\omega, k) = 1 - \omega_p^2/(\omega^2 - 3k^2 v_T^2)$, where $3v_T^2 \doteq \pd_{\bar{n}} \Pi(\bar{n})/m$ \cite[Chap.~3]{book:stix}, \Eq{eq:ndop0} is then easily recovered.

%%%%%%%%%%%%%%%%%%%%%%%%%%%%%%%%%%%%%%% 
\section{Formulas for $\boldsymbol{j}$ and $\boldsymbol{G_{1,2}}$}
\label{app:J}

The explicit formulas for the auxiliary functions introduced in the main text can be expressed in terms of the following special functions,
\begin{align}
\msf{K}(\zeta) & \doteq \int_0^{\pi/2}(1-\zeta \sin^2 \theta)^{-1/2} d\theta, \notag\\
\msf{E}(\zeta) & \doteq \int_0^{\pi/2}(1-\zeta \sin^2 \theta)^{1/2} d\theta, \notag\\
\msf{\Pi}(\vartheta, \zeta) & \doteq \int_0^{\pi/2}(1-\vartheta \sin^2 \theta)^{-1}(1-\zeta \sin^2 \theta)^{-1/2} d\theta,\notag
\end{align}
which are the complete elliptic integrals of the first, second, and third kind, respectively \cite{foot:math}. 

%--------------------------------------
\subsection{Normalized action}

The explicit formulas for the function $j$ are as follows. For trapped particles ($z \leqslant r < 1 + z$), one has
\begin{gather}
j = \frac{2}{\pi}\left[A^{\msf{K}}_{\msf{t}} \msf{K}(R_{\msf{t}}) + A^{\msf{E}}_{\msf{t}} \msf{E}(R_{\msf{t}}) + A^{\msf{\Pi}}_{\msf{t}}\msf{\Pi}(N_{\msf{t}}, R_{\msf{t}})\right],\\
A^{\msf{K}}_{\msf{t}} = - D^{1/2} \, \frac{2 r + 6 z - 1 - D}{8 z + 1 - D}, \\
A^{\msf{E}}_{\msf{t}} = D^{1/2}, \\
A^{\msf{\Pi}}_{\msf{t}} = D^{-1/2} \, \frac{2 r + 6 z - 1 - D}{8 z + 1 - D}, \\
R_{\msf{t}} = \frac{2 r + 6 z - 1 + D}{2D},\\
N_{\msf{t}} = \frac{8 z - 1 + D}{2D}.
\end{gather}
For passing particles ($r > 1 + z$), one has
\begin{gather}
j = \frac{2}{\pi} \left[A^{\msf{K}}_{\msf{p}} \msf{K}(R_{\msf{p}}) + A^{\msf{E}}_{\msf{p}} \msf{E}(R_{\msf{p}}) + A^{\msf{\Pi}}_{\msf{p}} \msf{\Pi} (N_{\msf{p}},R_{\msf{p}})\right],\\
A^{\msf{K}}_{\msf{p}} = \frac{\sqrt{2} (r - z - 1)}{\sqrt{ 2 r + 6 z - 1 + D}} \, \frac{8 z - 1 + D}{8 z + 1 + D},\\
A^{\msf{E}}_{\msf{p}} = \sqrt{\frac{2 r + 6 z - 1 + D}{2}},\\
A^{\msf{\Pi}}_{\msf{p}} = \frac{2 \sqrt{2}(r - z - 1)}{\sqrt{2 r + 6 z - 1 + D}} \, \frac{1}{8 z + 1 + D},\\
R_{\msf{p}} = \frac{2D}{2 r + 6 z - 1 + D},\\
N_{\msf{p}} = \frac{1 - 8z + D}{2(r - z)}.
\end{gather}

Note also that, for passing particles, $D$ is real for all $r$ when $z \geqslant 0$ and for $r \leqslant 1/2 + |z| + 1/(32|z|)$ when $z < 0$. Even when $D$ is imaginary, however, the above formulas still yield $j$ as a real function. In particular, the first three coefficients in \Eq{eq:jexp0} are real functions given by
\begin{gather}
I_1 = \frac{4}{\pi}\,\sqrt{r}\,\msf{E}(r^{-1}), 
\end{gather}
\begin{multline} 
I_2 = \frac{1}{3 \pi \sqrt{r}}\,[16 r (-1+2 r) \msf{E}(r^{-1})\\
      -2 \left(3-16 r+16 r^2\right) \msf{K}(r^{-1})], 
\end{multline}
\begin{multline}      
I_3 = \frac{1}{10 \pi (r - 1)\sqrt{r}}\\
      \times [(-5+224 r-1248 r^2+2048 r^3-1024 r^4) \msf{E}(r^{-1}) \\
      +16(5-47 r+138 r^2-160 r^3+64 r^4) \msf{K}(r^{-1})]. \notag
\end{multline}
It is to be noted that, in the limit $z \to 0$, these formulas are commonly known. For example, see \Ref{my:actii}. 

%--------------------------------------
\subsection{Functions $\boldsymbol{G_{1,2}}$}

The explicit formulas for the functions $G_{1,2}$ are as follows. For trapped particles ($z \leqslant r < 1 + z$), one has
\begin{gather}
G_1 = \bar{A}^{\msf{K}}_{\msf{t}} + \frac{\bar{A}^{\msf{\Pi}}_{\msf{t}} \msf{\Pi}(N_{\msf{t}}, R_{\msf{t}})}{\msf{K}(R_{\msf{t}})},\\
G_2 = \bar{B}^{\msf{K}}_{\msf{t}} + \frac{\bar{B}^{\msf{E}}_{\msf{t}} \msf{E}(R_{\msf{t}}) + \bar{B}^{\msf{\Pi}}_{\msf{t}} \msf{\Pi}(N_{\msf{t}},R_{\msf{t}})}{\msf{K}(R_{\msf{t}})},\\
\bar{A}^{\msf{K}}_{\msf{t}} = -\frac{1 + D}{8z},\\
\bar{A}^{\msf{\Pi}}_{\msf{t}} = \frac{1 - 8z + D}{8z},\\
\bar{B}^{\msf{K}}_{\msf{t}} = \frac{1 - 32z^2 +(1 + 8z)D}{32z^2},\\
\bar{B}^{\msf{E}}_{\msf{t}} = -\frac{D}{2z},\\  
\bar{B}^{\msf{\Pi}}_{\msf{t}} = -\frac{1 - 8z + D}{32z^2}.
\end{gather}
For passing particles ($r > 1 + z$), one has
\begin{gather}
G_1 = \bar{A}^{\msf{K}}_{\msf{p}} + \frac{\bar{A}^{\msf{\Pi}}_{\msf{p}} \msf{\Pi}(N_{\msf{p}}, R_{\msf{p}})}{\msf{K}(R_{\msf{p}})},\\
G_2 = \bar{B}^{\msf{K}}_{\msf{p}} + \frac{\bar{B}^{\msf{E}}_{\msf{p}} \msf{E}(R_{\msf{p}}) + \bar{B}^{\msf{\Pi}}_{\msf{p}}\msf{\Pi}(N_{\msf{p}}, R_{\msf{p}})}{\msf{K}(R_{\msf{p}})},\\
\bar{A}^{\msf{K}}_{\msf{p}} = -\frac{1 - D}{8z},\\
\bar{A}^{\msf{\Pi}}_{\msf{p}} = \frac{1 + 8z - D}{8z},\\
\bar{B}^{\msf{K}}_{\msf{p}} = \frac{1 - 8z + 16rz + 16z^2 - D}{32z^2},\\
\bar{B}^{\msf{E}}_{\msf{p}} = \frac{1 - 2r - 6z - D}{4z},\\
\bar{B}^{\msf{\Pi}}_{\msf{p}} = -\frac{1 + 8z - D}{32z^2}.
\end{gather}
It is to be noted that, in the limit $z \to 0$, the function $G_1$ was reported also in \Ref{my:actii}.

%%%%%%%%%%%%%%%%%%%%%%%%%%%%%%%%%%%%%%% 
\section{Approximations for integrals over smooth distributions}
\label{app:smooth}

Here, we show how to approximate integrals like
\begin{gather}
\mc{G} \doteq \int_0^\infty G F(J)\,dJ,
\end{gather}
where $F$ is smooth compared to $G$. We assume that $G \equiv G(r(j)) \equiv G(j)$ is either $G_1$ or $G_2$, so it has asymptotics [cf. \Eqs{eq:G1as} and \eq{eq:G2as}]
\begin{gather}\label{eq:Gasymgen}
G(j) = \frac{c_2}{j^2}+\frac{3c_4}{j^4}+\frac{5c_6}{j^6}+\ldots,
\end{gather}
where $c_q$ are independent of $j$. (The dependence on $z$ is implied but will not be emphasized in this appendix.) Then, an approximate formula for $\mc{G}$ is derived as follows. 

First of all, let us introduce an auxiliary function
\begin{gather}
Y(J) \doteq -\int_0^J G\,d\tilde{J}.
\end{gather}
It is convenient to express it also as $Y = \hat{J}\sqrt{a}\,\Psi$, where
\begin{gather}
\Psi \doteq -\int_0^j G (\tilde{\jmath})\,d\tilde{\jmath}
\end{gather}
is a dimensionless function with the following properties:
\begin{gather}\label{eq:Psiasym}
\Psi(j \to 0) = 0, \quad \Psi(j \to \infty) = 0.
\end{gather}
The former equality is obvious from the definition, and the latter equality is understood as follows. Notice that, with $G$ being either of $G_i$
($i = 1,2$), the function $\Psi(j)$ is proportional to $\mc{G}_i$ calculated on the waterbag (flat) distribution with thermal speed $\bar{v} \propto
j$. At $\bar{v} \to \infty$, the boundaries of this distribution are infinitely far from the resonance, so the trajectories on these boundaries
are unaffected by the wave. Thus, $\ffavr{\mc{E}}$ is insensitive to the wave amplitude, and thus both $\mc{G}_1$ and $\mc{G}_2$ are zero. [For $z = 0$, this is also shown in \Ref{my:actii} via a straightforward calculation.] Thus, in the limit of large $j$, the function $\Psi(j)$ must vanish, which proves the second part of \Eq{eq:Psiasym} and also leads to the following asymptotics at large $j$:
\begin{gather}
\Psi(j) = \frac{c_2}{j}+\frac{c_4}{j^3}+\frac{c_6}{j^5} + \ldots
\end{gather}

Hence one can rewrite $\mc{G}$ as follows,
\begin{gather}
\mc{G} = -\int_0^\infty Y'(J) F(J)\,dJ = \int_0^\infty Y(J) F'(J)\,dJ.
\end{gather}
Using the fact that $F'$ is smooth compared to $Y$, we can approximate the integrand, $Q(J) \doteq Y(J) F'(J)$, as
\begin{gather}\label{eq:Q}
Q(J) = Q_1(J) + Q_2(J) - Q_3(J),
\end{gather}
where the indexes denote the small-$J$, large-$J$, and intermediate-$J$ asymptotics, correspondingly. At small $J$, $Q_3$ cancels $Q_2$, so $Q(J)\approx Q_1(J)$. At large $J$, $Q_3$ cancels $Q_1$, so $Q(J)\approx Q_2(J)$. Thus, \Eq{eq:Q} approximates the true function $Q$ accurately at all $J$.

To construct the specific expressions for $Q_{1,3}$, notice that the derivatives of $F$ at $J = 0$, $F_0^{(q)}\doteq F^{(q)}(0)$, are yielded by \Eq{eq:Ff0} in the form \cite{foot:cont}
\begin{gather}
F_0^{(q)} = 2f_0^{(q)}(u)\left(\frac{k}{m}\right)^{\!q+1}\!\!\times 
\left\{
\begin{array}{ll}
 0, & q\text{ is odd}, \\
 1, & q\text{ is even}. \\
\end{array}
\right.
\end{gather}
Then, much like in \Ref{my:actii}, one can take
\begin{widetext}
\begin{gather}
Q_1(J) = Y(J)\bigg(F_0^{(2)}J+F_0^{(4)}\frac{J^3}{3!}+F_0^{(6)}\frac{J^5}{5!}+\ldots\bigg),\\
Q_2(J) = \hat{J}^2a\bigg(\frac{c_2}{J}+\frac{c_4}{J^3}\,\hat{J}^2a+\frac{c_6}{J^5}\,\hat{J}^4a^2+\ldots\bigg)F'(J),\\
Q_3(J) = \hat{J}^2a\bigg(\frac{c_2}{J}+\frac{c_4}{J^3}\,\hat{J}^2a+\frac{c_6}{J^5}\,\hat{J}^4a^2+\ldots\bigg)
\bigg(F_0^{(2)}J+F_0^{(4)}\frac{J^3}{3!}+F_0^{(6)}\frac{J^5}{5!}+\ldots\bigg).
\end{gather}

After substituting this into \Eq{eq:Q} and rearranging the terms, we can express $Q$ as
\begin{multline}\notag
Q =\hat{J}^2a\left[
F'(J) \left(\frac{c_2}{J} + \frac{c_4}{J^3}\,\hat{J}^2a + \frac{c_6}{J^5}\,\hat{J}^4a^2\right)
- F_0^{(2)}J \left(\frac{c_4}{J^3}\,\hat{J}^2a + \frac{c_6}{J^5}\,\hat{J}^4a^2\right)
- F_0^{(4)} \frac{J^3}{3!}\left(\frac{c_6}{J^5}\,\hat{J}^4a^2\right)
\right]\\
+ F_0^{(2)} (\hat{J}^2a)\, [\Psi(j)j - c_2] 
+ F_0^{(4)} (\hat{J}^2a)^2\,\frac{1}{3!}\,[\Psi(j)j^3 - c_2 j^2 - c_4]
+ F_0^{(6)} (\hat{J}^2a)^3\frac{1}{5!}\,[\Psi(j)j^5 - c_2 j^4 - c_4 j^2 - c_6]
+ \ldots
\end{multline}
Hence, we obtain
\begin{multline}
\mc{G} = \big[(\hat{J}^2a) c_2 \chi^{(2)} + (\hat{J}^2a)^2 c_4 \chi^{(4)} + (\hat{J}^2a)^3 c_6 \chi^{(6)} + \ldots\big]
\\+ \big[
  (\hat{J}^2a)^{3/2} F_0^{(2)} \eta^{(2)}
+ (\hat{J}^2a)^{5/2} F_0^{(4)} \eta^{(4)}
+ (\hat{J}^2a)^{7/2} F_0^{(6)} \eta^{(6)}
+ \ldots
\big],\label{eq:Gfirstexp}
\end{multline}
where the coefficients are defined as the following (converging) integrals:
\begin{gather}
\chi^{(2)} \doteq \int_0^\infty J^{-1} F'(J)\,dJ,\\
\chi^{(4)} \doteq \int_0^\infty J^{-3}[F'(J) - F_0^{(2)}J]\,dJ,\\
\chi^{(6)} \doteq \int_0^\infty J^{-5} [F'(J) - F_0^{(2)}J - F_0^{(4)} J^3/3!]\,dJ,\\
\eta^{(2)} \doteq \int_0^\infty [\Psi(j)j - c_2]\,dj,\\
\eta^{(4)} \doteq \frac{1}{3!} \int_0^\infty [\Psi(j)j^3 - c_2 j^2 - c_4]\,dj,\\
\eta^{(6)} \doteq \frac{1}{5!} \int_0^\infty [\Psi(j)j^5 - c_2 j^4 - c_4 j^2 - c_6]\,dj.
\end{gather}

It is convenient to express $\chi^{(2)}$ in terms of the dielectric function given by \Eq{eq:epsP}. This is done as follows:
\begin{align}
\chi^{(2)} 
& = \frac{k^2}{m^2} \int_0^\infty \left[f_0'(u + kJ/m)-f_0'(u - kJ/m)\right]\frac{dJ}{J} \notag \\
& = \frac{k^2}{m^2} \lim_{\varepsilon \to 0}\int_{\varepsilon}^\infty \left[f_0'(u+w)-f_0'(u-w)\right]\frac{dw}{w} \notag\\
& = \frac{k^2}{m^2}\, \msf{P}\int_{-\infty}^\infty \frac{f_0'(v)}{v-u}\,dw \notag\\
& = \frac{k^4}{m^2\omega_p^2}\,[1-\epsilon (\omega, k)].
\end{align}
Using the formulas derived in \App{app:pv} [namely, \Eqs{eq:appd2} and \eq{eq:appd4}], we also obtain
\begin{align}
\chi^{(4)}
& = \frac{k^2}{m^2}\int_0^\infty [f_0'(u + kJ/m)-f_0'(u - kJ/m)-2f_0^{(2)}(u)\,(kJ/m)]\,\frac{dJ}{J^3} \notag\\
& = \frac{k^4}{m^4}\lim_{\varepsilon \to 0} \int_{\varepsilon}^\infty [f_0'(u+w)-f_0'(u-w)-2f_0^{(2)}(u)w]\,\frac{dw}{w^3} \notag\\
& = \frac{k^4}{m^4}\,\msf{P}\int_{-\infty}^\infty [f_0'(u+w)-f_0^{(2)}(u)w]\,\frac{dw}{w^3} \notag\\
& = \frac{k^4}{m^4}\,\frac{1}{2}\,\frac{\pd^2}{\pd u^2}\,\msf{P}\int_{-\infty}^\infty \frac{f_0'(v)}{v-u}\,dv \notag\\
& = -\frac{1}{2}\,\frac{k^8}{m^4\omega_p^2}\,\frac{\pd^2\epsilon(\omega, k)}{\pd \omega^2},
\end{align}
\begin{align}
\chi^{(6)}
& = \frac{k^2}{m^2}\int_0^\infty \left[
   f_0'(u+kJ/m)-f_0'(u-kJ/m)-2f_0^{(2)}(u)\,kJ/m-2f_0^{(4)}(u)\,\frac{(kJ/m)^3}{3!}
   \right]\frac{dJ}{J^5} \notag\\
& =\frac{k^6}{m^6}\lim_{\varepsilon \to 0}
   \int_{\varepsilon}^\infty \left[f_0'(u+w)-f_0'(u-w)-2f_0^{(2)}(u)w-2f_0^{(4)}(u)\, \frac{w^3}{3!}\right]\frac{dw}{w^5} \notag\\
& =\frac{k^6}{m^6}\,\msf{P}\int_{-\infty}^\infty \left[f_0'(u+w)-f_0^{(2)}(u)w-f_0^{(4)}(u)\, \frac{w^3}{3!}\right]\frac{dw}{w^5} \notag\\
& =\frac{k^6}{m^6}\frac{1}{4!}\,\frac{\pd^4}{\pd u^4}\,\msf{P}\int_{-\infty}^\infty \frac{f_0'(v)}{v-u}\,dv \notag\\
& =-\frac{1}{4!}\,\frac{k^{12}}{m^6\omega_p^2}\,\frac{\pd^4\epsilon (\omega, k)}{\pd \omega^4}.
\end{align}
Then one can rewrite \Eq{eq:Gfirstexp} as
\begin{multline}\label{eq:mcGfin}
\mc{G}
= \frac{a\omega^2}{\omega_p^2}\,\bigg\{[1-\epsilon(\omega, k)] c_2 
- \frac{\omega^2}{2!}\,\frac{\pd^2\epsilon(\omega, k)}{\pd \omega^2}\,c_4 a
- \frac{\omega^4}{4!}\,\frac{\pd^4\epsilon (\omega, k)}{\pd \omega^4}\,c_6 a^2 + \ldots \bigg\}\\
+ 2a\bigg\{\eta^{(2)} a^{1/2} u^3 f_0^{(2)}(u) + \eta^{(4)} a^{3/2} u^5 f_0^{(4)}(u) + \eta^{(6)} a^{5/2} u^7 f_0^{(6)}(u) + \ldots\bigg\}.
\end{multline}
\end{widetext}

Finally, notice the following. In \Eq{eq:mcGfin}, the term in the first curly brackets, which we call the fluid term, contains an expansion in integer powers of $a$. If $c_q$ with neighboring $q$ are about the same, the ratio of the neighboring terms is of the order of $a$. The term in the second curly brackets, which we call the kinetic term, contains an expansion in half-integer powers of $a$. If $\eta^{(q)}$ with neighboring $q$ are about the same, the ratio of the neighboring terms is about $a u^4/v_T^4$ (assuming, \eg the Maxwellian $f_0$), where $v_T$ is the thermal velocity. Hence, the applicability conditions of \Eq{eq:mcGfin} are
\begin{gather}\label{eq:appappl}
a \ll 1, \quad a u^4/v_T^4 \ll 1.
\end{gather}

That said, the applicability conditions may not actually be as strict as this formal estimate suggests. That is because, when the fluid term is much bigger than the kinetic term, the latter does not need to be calculated with high accuracy, and vice versa. It is also to be noted that, when deriving the conditions \eq{eq:appappl}, we entirely neglected order-one numerical coefficients. In practice, these coefficients seem to make higher-order terms less important. We conclude this from the results reported in \Sec{sec:smooth}, where our asymptotic formulas demonstrate reasonable agreement with more precise calculation based on the original NDR, \Eq{eq:ndr2}, even though the applicability conditions of the asymptotic theory are satisfied for the chosen parameters only marginally.

%%%%%%%%%%%%%%%%%%%%%%%%%%%%%%%%%%%%%%% 
\section{Derivatives of principal value integrals}
\label{app:pv}

In this appendix, we report some useful identities involving derivatives of principal-value integrals. It is likely that these formulas can
be found in the existing literature, but it is instructive to rederive them here. Specifically, we will consider integrals of the type
\begin{gather}
\msf{P}\int_{-\infty}^\infty \frac{h(v)}{v-u}\,dv \equiv \lim_{\varepsilon \to 0}\int_\varepsilon \frac{h(v)}{v-u}\,dv,
\end{gather}
where $h$ is some function, and where the following notation is introduced,
\begin{gather}
\int_\varepsilon (\ldots)\,dv \doteq 
\int_{-\infty}^{u-\varepsilon} (\ldots)\,dv + \int_{u+\varepsilon}^\infty (\ldots)\,dv.
\end{gather}
Note one special type of such integrals,
\begin{gather}\label{eq:epspv}
(2\ell-1)\int_\varepsilon  \frac{1}{(v-u)^{2\ell}}\,dv = \frac{2}{\varepsilon^{2\ell-1}}, \quad \ell \in \mathbb{N},
\end{gather}
which is a convenient formula to be used below.

%--------------------------------------
\subsection{Second derivative}
\label{sec:pv1}

Consider an expression of the form
\begin{widetext}
\begin{align}
P_2[h] 
& \doteq \frac{1}{2}\,\frac{\pd^2}{\pd u^2}\,\msf{P}\int_{-\infty}^\infty \frac{h(v)}{v-u}\,dv \notag\\
& = \frac{1}{2}\,\frac{\pd^2}{\pd u^2}\,\lim_{\varepsilon \to 0} 
    \left[
    \int_{-\infty}^{u-\varepsilon} \frac{h(v)}{v-u}\,dv + \int_{u+\varepsilon}^\infty \frac{h(v)}{v-u}\,dv
    \right]\notag\\
& = \frac{1}{2}\,\frac{\pd}{\pd u}\,\lim_{\varepsilon \to 0} 
    \left[
    \frac{h(u-\varepsilon)}{-\varepsilon} + \int_{-\infty}^{u-\varepsilon} \frac{h(v)}{(v-u)^2}\,dv 
    + \int_{u+\varepsilon}^\infty \frac{h(v)}{(v-u)^2}\,dv
    - \frac{h(u+\varepsilon)}{\varepsilon}
    \right]\notag\\
& = \lim_{\varepsilon \to 0} \left[
    \frac{h'(u-\varepsilon)}{-2\varepsilon} + \frac{h(u-\varepsilon)}{2\varepsilon^2}
    + \int_\varepsilon \frac{h(v)}{(v-u)^3}\,dv 
    - \frac{h(u+\varepsilon)}{2\varepsilon^2} - \frac{h'(u+\varepsilon)}{2\varepsilon}
    \right]\notag\\
& = \lim_{\varepsilon \to 0} \left[ \int_\varepsilon \frac{h(v)}{(v-u)^3}dv + \Delta P_2\right].
\end{align}
Here we introduced
\begin{gather}
\Delta P_2 \doteq 
  - \frac{h'(u-\varepsilon)}{2\varepsilon} 
  + \frac{h(u-\varepsilon)}{2\varepsilon^2} - \frac{h(u+\varepsilon)}{2\varepsilon^2} - \frac{h'(u+\varepsilon)}{2\varepsilon}
= - \frac{2h'(u)}{\varepsilon} + o(1)
= - \int_\varepsilon \frac{h'(u)}{(v-u)^2}\,dv + o(1),
\end{gather}
where \Eq{eq:epspv} was substituted. Hence,
\begin{gather}\label{eq:pvp2}
P_2[h] 
= \lim_{\varepsilon \to 0} \left[\int_\varepsilon \frac{h(v)}{(v-u)^3}\,dv - \int_\varepsilon \frac{h'(u)}{(v-u)^2}\,dv\right]
= \msf{P} \int_{-\infty}^\infty \left[\frac{h(v)}{(v-u)^3} - \frac{h'(u)}{(v-u)^2}\right]dv.
\end{gather}
This leads to the main result of this subsection, which is that
\begin{gather}\label{eq:appd2}
\frac{1}{2!}\frac{\pd^2}{\pd u^2}\,\msf{P}\int_{-\infty}^\infty \frac{h(v)}{v-u}\,dv
= \msf{P}\int_{-\infty}^\infty \frac{h(v) - h'(u)(v-u)}{(v-u)^3}\,v.
\end{gather}

%--------------------------------------
\subsection{Fourth derivative}

The fourth derivative is treated similarly:
\begin{align}
P_4[h]
& \doteq \frac{1}{4!}\,\frac{\pd^4}{\pd u^4}\,\msf{P} \int_{-\infty}^\infty \frac{h(v)}{v-u}\,dv \notag\\
& = \frac{2}{4!}\,\frac{\pd^2}{\pd u^2}P_2[h] \notag\\
& = \frac{1}{12}\,\frac{\pd^2}{\pd u^2}\,\msf{P}\int_{-\infty}^\infty \left[\frac{h(v)}{(v-u)^3}-\frac{h'(u)}{(v-u)^2}\right] dv, \notag \\
& = \frac{1}{12}\,\frac{\pd}{\pd u}\,\lim_{\varepsilon \to 0}
    \left\{
     \int_\varepsilon 
       \left[\frac{3h(v)}{(v-u)^4} - \frac{2h'(u)}{(v-u)^3} - \frac{h''(u)}{(v-u)^2}\right]dv
    +\Delta P_4    
    \right\}
    \notag\\
& = \frac{1}{12}\,\lim_{\varepsilon \to 0}
    \left\{
     \frac{\pd}{\pd u}\int_\varepsilon \left[\frac{3h(v)}{(v-u)^4}-\frac{h''(u)}{(v-u)^2}\right]dv
     +\frac{\pd}{\pd u}\,\Delta P_4
    \right\}
\end{align}
[the term $2h'(u)(v-u)^{-3}$ was dropped because the principal value integral over it is zero], where
\begin{gather}
\Delta P_4 \doteq 
\left[\frac{h(v)}{(v-u)^3} - \frac{h'(u)}{(v-u)^2}\right]_{v = u-\varepsilon} 
- \left[\frac{h(v)}{(v-u)^3}-\frac{h'(u)}{(v-u)^2}\right]_{v = u+\varepsilon} 
= - \frac{1}{\varepsilon^3}\,[h(u-\varepsilon) + h(u+\varepsilon)].
\end{gather}
Then,
\begin{gather}\notag
P_4[h] 
= \frac{1}{12}\,\lim_{\varepsilon \to 0}
 \left\{
 \frac{\pd}{\pd u} \int_\varepsilon \left[\frac{3h(v)}{(v-u)^4} - \frac{h''(u)}{(v-u)^2}\right]dv
 + \frac{\pd}{\pd u}\,\Delta  P_4
 \right\}
= \lim_{\varepsilon \to 0}
 \left\{
 \int_\varepsilon \left[\frac{h(v)}{(v-u)^5}-\frac{h'''(u)}{12(v-u)^2}\right]dv + \frac{\Delta R_4}{12}
 \right\},
\end{gather}
where we introduced
\begin{gather} 
\Delta R_4 
\doteq \left[
  \frac{3h(v)}{(v-u)^4} - \frac{h''(u)}{(v-u)^2}\right]_{v = u-\varepsilon}
  - \left[\frac{3h(v)}{(v-u)^4} - \frac{h''(u)}{(v-u)^2}\right]_{v = u+\varepsilon}
  +\frac{\pd}{\pd u}\, \Delta P_4.
\end{gather}
It is easy to see that
\begin{gather}  \notag
\Delta R_4  = \frac{3}{\varepsilon^4}\,[h(u-\varepsilon)-h(u+\varepsilon)]- \frac{1}{\varepsilon^3}\,[h'(u-\varepsilon) + h'(u+\varepsilon)]
= - \frac{8h'(u)}{\varepsilon^3} - \frac{2h'''(u)}{\varepsilon}
= \int_\varepsilon \left[-\frac{12h'(u)}{(v-u)^4} - \frac{h'''(u)}{(v-u)^2}\right]dv,
\end{gather}
where we substituted \Eq{eq:epspv}. Reverting to $P_2$, we then obtain
\begin{gather}
P_4[h] = \msf{P}\int_{-\infty}^\infty \left[\frac{h(v)}{(v-u)^5} - \frac{h'(u)}{(v-u)^4} - \frac{h'''(u)}{6(v-u)^2}\right] dv.
\end{gather}
This leads to the main result of this subsection, which is that
\begin{gather}\label{eq:appd4}
\frac{1}{4!}\,\frac{\pd^4}{\pd u^4}\,\msf{P}\int_{-\infty}^\infty \frac{h(v)}{v-u}\,dv
= \msf{P}\int_{-\infty}^\infty \frac{1}{(v-u)^5} \left[h(v)- h'(u) (v-u) - h'''(u)\,\frac{(v-u)^3}{3!}\right] dv.
\end{gather}

\end{widetext}


\begin{thebibliography}{10}

\bibitem{ref:akhiezer56}
A.~I. Akhiezer and R.~V. Polovin, Zh. Eksp. Teor. Fiz. {\bf 30}, 915 (1956)
  [Sov. Phys. JETP {\bf 3}, 696 (1956)].

\bibitem{ref:tidman65}
D.~A. Tidman and H.~M. Stainer, Phys. Fluids {\bf 8}, 345 (1965).

\bibitem{ref:bertrand69}
P.~Bertrand, G.~Baumann, and M.~R. Feix, Phys. Lett. {\bf 29A}, 489 (1969).

\bibitem{ref:coffey71}
T.~P. Coffey, Phys. Fluids {\bf 14}, 1402 (1971).

\bibitem{ref:dewar72e}
R.~L. Dewar and J.~Lindl, Phys. Fluids {\bf 15}, 820 (1972).

\bibitem{ref:mckinstrie87}
C.~J. McKinstrie and D.~W. Forslund, Phys. Fluids {\bf 30}, 904 (1987).

\bibitem{ref:manheimer71a}
W.~M. Manheimer and R.~W. Flynn, Phys. Fluids {\bf 14}, 2393 (1971).

\bibitem{ref:morales72}
G.~J. Morales and T.~M. O'Neil, Phys. Rev. Lett. {\bf 28}, 417 (1972).

\bibitem{ref:lee72}
A.~Lee and G.~Pocobelli, Phys. Fluids {\bf 15}, 2351 (1972).

\bibitem{ref:dewar72b}
R.~L. Dewar, Phys. Fluids {\bf 15}, 712 (1972).

\bibitem{ref:kim76}
H.~Kim, Phys. Fluids {\bf 19}, 1362 (1976).

\bibitem{ref:rose01}
H.~A. Rose and D.~A. Russell, Phys. Plasmas {\bf 8}, 4784 (2001).

\bibitem{ref:barnes04}
D.~C. Barnes, Phys. Plasmas {\bf 11}, 903 (2004).

\bibitem{ref:rose05}
H.~A. Rose, Phys. Plasmas {\bf 12}, 012318 (2005).

\bibitem{ref:lindberg07}
R.~R. Lindberg, A.~E. Charman, and J.~S. Wurtele, Phys. Plasmas {\bf 14},
  122103 (2007).

\bibitem{ref:khain07}
P.~Khain and L.~Friedland, Phys. Plasmas {\bf 14}, 082110 (2007).

\bibitem{ref:benisti07}
D.~B\'enisti and L.~Gremillet, Phys. Plasmas {\bf 14}, 042304 (2007).

\bibitem{ref:benisti08}
D.~B\'enisti, D.~J. Strozzi, and L.~Gremillet, Phys. Plasmas {\bf 15}, 030701
  (2008).

\bibitem{ref:benisti09}
D.~B\'enisti, D.~J. Strozzi, L.~Gremillet, and O.~Morice, Phys. Rev. Lett. {\bf
  103}, 155002 (2009).

\bibitem{ref:matveev09}
A.~I. Matveev, Rus. Phys. J. {\bf 52}, 885 (2009).

\bibitem{ref:schamel12}
H.~Schamel, Phys. Plasmas {\bf 19}, 020501 (2012).

\bibitem{ref:goldman71}
M.~V. Goldman and H.~L. Berk, Phys. Fluids {\bf 14}, 801 (1971).

\bibitem{ref:krasovsky94}
V.~L. Krasovsky, Physica Scripta {\bf 49}, 489 (1994).

\bibitem{ref:krasovsky92}
V.~L. Krasovsky, Phys. Lett. A {\bf 163}, 199 (1992).

\bibitem{ref:krasovskii95}
V.~L. Krasovskii, Zh. Eksp. Teor. Fiz. {\bf 107}, 741 (1995) [JETP {\bf 80},
  420 (1995)].

\bibitem{ref:krasovsky07}
V.~L. Krasovsky, J. Plasma Phys. {\bf 73}, 179 (2007).

\bibitem{ref:breizman10}
B.~N. Breizman, Nucl. Fusion {\bf 50}, 084014 (2010).

\bibitem{my:trcomp}
P.~F. Schmit, I.~Y. Dodin, J.~Rocks, and N.~J. Fisch, Phys. Rev. Lett. {\bf
  110}, 055001 (2013).

\bibitem{foot:berger13}
For a literature review and a relevant recent study, see R.~L. Berger,
  S.~Brunner, T.~Chapman, L.~Divol, C.~H. Still, and E.~J. Valeo, Phys. Plasmas
  {\bf 20}, 032107 (2013).

\bibitem{foot:breizman11}
For a review, see B.~Breizman, Fusion Sci. Tech. {\bf 59}, 549 (2011).

\bibitem{ref:winjum07}
B.~J. Winjum, J.~Fahlen, and W.~B. Mori, Phys. Plasmas {\bf 14}, 102104 (2007).

\bibitem{ref:bernstein57}
I.~B. Bernstein, J.~M. Greene, and M.~D. Kruskal, Phys. Rev. {\bf 108}, 546
  (1957).

\bibitem{my:itervar}
I.~Y. Dodin, Fusion Sci. Tech. {\bf 65}, 54 (2014).

\bibitem{my:bgk}
I.~Y. Dodin and N.~J. Fisch, Phys. Rev. Lett. {\bf 107}, 035005 (2011).

\bibitem{my:acti}
I.~Y. Dodin and N.~J. Fisch, Phys. Plasmas {\bf 19}, 012102 (2012).

\bibitem{my:actii}
I.~Y. Dodin and N.~J. Fisch, Phys. Plasmas {\bf 19}, 012103 (2012).

\bibitem{my:actiii}
I.~Y. Dodin and N.~J. Fisch, Phys. Plasmas {\bf 19}, 012104 (2012).

\bibitem{book:whitham}
G.~B. Whitham, {\it Linear and Nonlinear Waves\/} (Wiley, New York, 1974).

\bibitem{foot:khain}
A related Lagrangian formulation was proposed recently also in P.~Khain and
  L.~Friedland, Phys. Plasmas {\bf 17}, 102308 (2010); P.~Khain, L.~Friedland,
  A.~G. Shagalov, and J.~S. Wurtele, Phys. Plasmas {\bf 19}, 072319 (2012).

\bibitem{foot:diss}
For a discussion regarding dissipation, see, \eg D.~B\'enisti, O.~Morice, and
  L.~Gremillet, Phys. Plasmas {\bf 19}, 063110 (2012).

\bibitem{foot:static}
This is because nonzero $\bar{E}$ makes the passing-particle dynamics aperiodic
  irrespective of how smooth the wave field is. An exception is the case when
  the slow field is weak enough such that, locally, $\bar{E}$-driven
  perturbations to the particle oscillations can be neglected. In that case, a
  simpler approach to accommodate $\bar{E}$ is possible, as discussed in
  Sec.~IV\,D of \Ref{my:itervar}.

\bibitem{foot:clear}
As will also be restated in \Sec{sec:gdr}, the NDR flows from the requirement
  that the derivatives of $\mcc{L}$ with respect to certain measures of the
  wave amplitude must be zero.

\bibitem{foot:benisti14}
For instance, see \Ref{ref:lindberg07} and D.~B\'{e}nisti, A.~Friou, and
  L.~Gremillet, Interdiscip. J. Discontin. Nonlinearity Complex. {\bf 3}, 435
  (2014).

\bibitem{book:stix}
T.~H. Stix, {\it Waves in Plasmas\/} (AIP, New York, 1992).

\bibitem{my:nmi}
I.~Y. Dodin, P.~F. Schmit, J.~Rocks, and N.~J. Fisch, Phys. Rev. Lett. {\bf
  110}, 215006 (2013).

\bibitem{foot:addif}
The effect is related to what is sometimes called ``adiabatic diffusion'' in
  quasilinear theory \cite[Chap.~16]{book:stix}. Also notably, the fact that
  wave frame is not inertial is pointed out in \Ref{ref:benisti08}.

\bibitem{my:amc}
I.~Y. Dodin and N.~J. Fisch, Phys. Rev. A {\bf 86}, 053834 (2012).

\bibitem{foot:sagdeev}
For instance, see R.~Z. Sagdeev, D.~A. Usikov, and G.~M. Zaslavsky, {\it
  Nonlinear Physics: from the Pendulum to Turbulence and Chaos\/} (Harwood
  Academic Publishers, New York, 1988), pp.~23-25.

\bibitem{foot:keen}
An exceptions to this rule are the so-called kinetic electrostatic electron
  nonlinear (KEEN) waves. For instance, see I.~Y. Dodin and N.~J. Fisch, Phys.
  Plasmas {\bf 21}, 034501 (2014) and the references cited therein.

\bibitem{foot:dJ}
At fixed $J$ (as well as $a$, $z$, $\omega$, and $k$), one has $0 = dJ =
  (\pd_\chi J)d\chi + (\pd_r J)dr$, so $\pd_\chi r(J, a, z, \omega, k) =
  -(\pd_\chi J)/(\pd_r J)$.

\bibitem{foot:sym}
The fact that the steady-state wave potential is an even function of the phase
  with respect to its extrema is seen, for instance, in \Ref{my:trcomp} or
  T.~W. Johnston, Y.~Tyshetskiy, A.~Ghizzo, and P.~Bertrand, Phys. Plasmas {\bf
  16}, 042105 (2009).

\bibitem{foot:z}
Assuming that $z$ is small enough, it is sufficient to replace $J$ with its
  linear expansion in $z$, using that $\pd_z J$ is continuous (see below). To
  do this, let us write $z \cos(2\theta + \chi) = z_1 \cos(2\theta) + z_2
  \sin(2\theta)$, where $z_1 \doteq z\cos(\chi)$ and $z_2 \doteq -z\sin(\chi)$
  can be used as independent variables instead of $(z, \chi)$. Then, $J \approx
  J_0 + (\pd_{z_1} J)_0 z_1 + (\pd_{z_2} J)_0 z_2$, where the index 0 denotes
  that the functions are evaluated at $z = 0$. But it is easy to see that
  $(\pd_{z_2} J)_0 = 0$, so $J \approx J_0 + (\pd_{z_1} J)_0 z_1$. The
  $\chi$-dependent part of $J$ is then proportional to $\cos(\chi)$, so it has
  extrema only at $\chi = 0$ and $\chi = \pi$.

\bibitem{foot:math}
The calculations reported in this paper were facilitated by
  \textsl{Mathematica} \textcopyright\ 1988-2011 Wolfram Research, Inc.,
  version number 8.0.4.0.

\bibitem{foot:waterbag}
For an alternative, hydrodynamic treatment of the waterbag model, see, \eg
  P.~Bertrand and M.~R. Feix, Phys. Lett. {\bf 28A}, 68 (1968).

\bibitem{foot:eaw}
In addition to Langmuir waves, the distributions \eq{eq:kappadistr} and
  \eq{eq:maxwellf} also support electron acoustic waves. See J.~P. Holloway and
  J.~J. Dorning, Phys. Rev. A {\bf 44}, 3856 (1991); F.~Valentini, T.~M.
  O'Neil, and D.~H.~E. Dubin, Phys. Plasmas {\bf 13}, 052303 (2006).

\bibitem{ref:livadiotis13}
G.~Livadiotis and D.~J. McComas, Space Sci. Rev. {\bf 175}, 183 (2013).

\bibitem{ref:friedland06}
L.~Friedland, P.~Khain, and A.~G. Shagalov, Phys. Rev. Lett. {\bf 96}, 225001
  (2006).

\bibitem{foot:calc}
From the Fourier-transformed Gauss's law, one has $k^2\epsilon(\omega,
  k)\varphi_1 = - 8 \pi e \bar{n}_{\msf{t}}$ and $(2k)^2\epsilon(2\omega,
  2k)\varphi_2 = 8 \pi e \bar{n}_{\msf{t}}$, where we used that the spatial
  profile of the trapped particle density is delta-shaped. Combining this with
  \Eq{eq:eps0} for $\epsilon$ and with \Eq{eq:linom} for $\omega_0$ leads to
  \Eqs{eq:aux48}.

\bibitem{foot:brz}
Equations \eq{eq:brz} and \eq{eq:brw} were obtained via Fourier-transforming
  the solution from \Ref{ref:breizman10}. Note that $\mc{N}$ is defined in
  terms of $a_p \propto \phi_1$ rather than in terms of the peak magnitude of
  the wave potential.

\bibitem{foot:eremin}
For a detailed study of this effect see D.~Yu. Eremin and H.~L. Berk, Phys.
  Plasmas {\bf 11}, 3621 (2004).

\bibitem{foot:period}
Nevertheless, the model used in \Refs{ref:mckinstrie87, ref:lindberg07} can be
  justified when boundary conditions are adopted to model a periodic
  \textit{piece} of otherwise aperiodic plasma. The relation between those
  theories and ours is discussed in \Sec{sec:doppler}.

\bibitem{foot:qin}
H.~Qin, J.~W. Burby, and R.~C. Davidson, Phys. Rev. E {\bf 90}, 043102 (2014).
  Similar results are obtained also using a continuous field theory derived
  independently from the semiclassical approximation (to be published).

\bibitem{foot:cont}
Here we assume, for simplicity, that all $f_0^{(q)}(V_0)$ of interest are
  continuous at $V_0 = u$. A more general case, leading to logarithmic
  nonlinearities, can be treated similarly to how the zero-$z$ case was treated
  in \Ref{my:actii}.

\end{thebibliography}
\end{document}